\documentclass[12pt]{iopart}

\usepackage{graphicx}
\usepackage{iopams}
\usepackage{xcolor}
\usepackage{multirow}
\usepackage{booktabs}
\usepackage{hyperref}
\hypersetup{
    colorlinks=true,
    linkcolor=black,
    filecolor=magenta,      
    urlcolor=blue,
    citecolor=blue,
    }
\usepackage[style=numeric, sorting=none, style=phys]{biblatex}
\bibliography{main.bib}
\begin{document}

\title[]{Towards a Machine-Learned Poisson Solver for Low-Temperature Plasma Simulations in Complex Geometries}

\author{Ihda Chaerony Siffa$^{1, 2, *}$, Markus M. Becker$^{1}$, Klaus-Dieter Weltmann$^{1}$, and Jan Trieschmann$^{2}$}

\address{$^{1}$ Leibniz Institute for Plasma Science and Technology (INP), Felix-Hausdorff-Straße 2, 17489 Greifswald, Germany}
\address{$^{2}$ Department of Electrical and Information Engineering, Kiel University, Kaiserstraße 2, 24143 Kiel, Germany}
\ead{*ihda.chaeronysiffa@inp-greifswald.de}
\vspace{10pt}
\begin{indented}
\item[]31 May 2023
\end{indented}

\begin{abstract}
Poisson's equation plays an important role in modeling many physical systems. In electrostatic self-consistent low-temperature plasma (LTP) simulations, Poisson's equation is solved at each simulation time step, which can amount to a significant computational cost for the entire simulation. In this paper, we describe the development of a generic machine-learned Poisson solver specifically designed for the requirements of LTP simulations in complex 2D reactor geometries on structured Cartesian grids. Here, the reactor geometries can consist of inner electrodes and dielectric materials as often found in LTP simulations. The approach leverages a hybrid CNN-transformer network architecture in combination with a weighted multiterm loss function. We train the network using highly-randomized synthetic data to ensure the generalizability of the learned solver to unseen reactor geometries. The results demonstrate that the learned solver is able to produce quantitatively and qualitatively accurate solutions. Furthermore, it generalizes well on new reactor geometries such as reference geometries found in the literature. To increase the numerical accuracy of the solutions required in LTP simulations, we employ a conventional iterative solver to refine the raw predictions, especially to recover the high-frequency features not resolved by the initial prediction. With this, the proposed learned Poisson solver provides the required accuracy and is potentially faster than a pure GPU-based conventional iterative solver. This opens up new possibilities for developing a generic and high-performing learned Poisson solver for LTP systems in complex geometries.

%With this, the proposed learned Poisson solver provides the required accuracy and is still considerably faster than traditional solvers, i.e. it is attractive for the simulation of LTP systems in complex geometries.

%To increase the accuracy of the solutions, we employ a conventional solver to refine the raw predictions, especially to recover the high-frequency features unresolved in the initial prediction by the learned solver. Using this approach, one can achieve reasonable solution accuracy that is still considerably faster than using a GPU-based conventional solver alone, which is an attractive alternative for simulating LTP systems in complex geometries.
\end{abstract}

%\submitto{\JPD}
%\maketitle

\section{Introduction}
%\begin{itemize}
%    \item Introduction on the Poisson Equation and its application in computational plasma physics (LTP)
%    \item Argument why it is interesting to use a data-driven approach to make a more efficient Poisson solver
%    \item A paragraph about related works
%\end{itemize}
Poisson's equation is an elliptic partial differential equation (PDE) of great relevance. It is used in the theoretical description of many physical systems, such as hydrodynamics, Newtonian gravity, quantum mechanics, and electrostatics/magnetostatics. For example, Poisson's equation describes a gravitational field generated by a massive object in a Newtonian system \cite{Low-1997-ID6172}. In a hydrodynamic system of incompressible flows, one solves the equation to obtain the scalar pressure field \cite{Wesseling-2009-ID6173}. In this work, we are particularly interested in the use of Poisson's equation for the modeling of low-temperature plasma (LTP) systems. Low-temperature plasmas are one example where the electrostatic approximation is frequently applied \cite{Alves-2018-ID6174}. Therein the electric potential across a domain arises from electric charges within the domain and the domain boundaries. Moreover, they present physical systems, where the dynamics of the electromagnetic fields and the charged species are strongly coupled. The numerical simulation of such coupled systems requires a solution of Poisson's equation in every simulation time or iteration step. The obtained potential distribution at the current time step is used to calculate the electric field, which directly influences the dynamics of the charged species, thus the outcome of the plasma simulation. Notice that although the subsequent discussion is centered around the electrostatic approximation, it can be easily transferred to the magnetostatic approximation or to electrodynamics using the Helmholtz equation \cite{Jackson-1998-ID6175}.

The repeated calculations for solving Poisson's equation over numerous simulation time steps amount to a significant computational cost of the entire simulation. This is especially true in multi-dimensional simulation settings, where the time for solving the equation represents a large part of the overall time consumption. Many numerical methods have been developed for solving elliptic boundary value problems (BVPs) such as the Poisson equation. These include the multigrid methods, which are currently among the most efficient iterative numerical solvers \cite{Hackbusch-1985-ID6176,Trottenberg-2001-ID6177,Teunissen-2023-ID6136}. Recently, a spectral element method based on the hierarchical Poincar\'e-Steklov scheme has shown promising results in terms of fast convergence and computational efficiency \cite{Martinsson-2013-ID6178,Fortunato-2021-ID6179,Semenov-2022-ID6018}. Nevertheless, the elliptic nature of Poisson's equation implies information propagation throughout the complete domain, limiting possible performance optimization techniques of the applied numerical methods \cite{Semenov-2022-ID6018}.

In the field of machine learning, artificial neural networks (ANNs) have been continuously improved in solving real-world tasks (e.g., image classifications, object recognitions, machine translations, etc.) since its breakthroughs in the early and mid-2010s, particularly through the use of modern ANN architectures such as convolutional neural networks (CNNs) \cite{Krizhevsky-2012-ID6180}, and the invention of the transformer networks \cite{Vaswani-2017-ID6181}. These breakthroughs were facilitated by advances in graphics processing units (GPUs) that have made training deep networks on large datasets computationally feasible \cite{Raina-2009-ID6182,ClaudiuCiresan-2010-ID6183,Krizhevsky-2012-ID6180}, in addition to the increasing availability of data required for the training.

The application of ANNs for solving PDEs dates back at least to the 1990s \cite{Lee-1990-ID6184,Dissanayake-1994-ID6185,Lagaris-1998-ID6186}. Due to the aforementioned recent successes of ANNs, coupled with the ever-advancing computing hardware, the research interest in solving PDEs with ANNs has gained further momentum. Speedup in computation time and generalization of ill-posed problems remain major motivations. Among the influential works in this area are \textit{physics-informed neural networks} (PINNs) \cite{Raissi-2019-ID6187,Karniadakis-1970-ID6188,Lu-2021-ID6189} and \textit{neural operators} (NOs) \cite{Lu-2019-ID6190, Li-2020-ID6191, Li-2020-ID6192, Cao-2021-ID6193, Lu-2021-ID6194}. In \cite{Shi-2022-ID6195}, Shi \textit{et al}. have proposed a data-free paradigm for NOs, and Wang \textit{et al.} have combined ideas from PINNs and NOs for a more data-efficient operator learning in \cite{Wang-2021-ID6196}.

It is worth noting that solving PDEs using ANNs has sound theoretical foundations, e.g., Hornik \textit{et al.} \cite{Hornik-1989-ID6197,Hornik-1991-ID6198} have shown that ANNs with multiple layers can approximate any continuous functions, and recently Zhou \textit{et al.} \cite{Zhou-2020-ID6199} have extended the earlier work to CNNs. In addition to the direct solution of PDEs using ANNs, \textit{neural solvers} with convergence guarantees have been proposed for fast iterative solutions \cite{Hsieh-2019-ID6200}.

On a more specific note, several recent works have applied CNN-based methods for solving \textit{specifically} the Poisson equation in spatially two-dimensional (2D) domains for fluid dynamic and electrostatic cases \cite{Ozbay-2021-ID6201,Zhang-2019-ID6202,Cheng-2021-ID6203}. These approaches aim to solve Poisson's equation subject to different source term values, boundary conditions, and domain sizes, without having to retrain the network. %This is unlike NOs, which often only use a coefficient as an input to predict the solution, and the training data is usually generated from one PDE with fixed boundary conditions.

In this paper, we propose a machine learning approach to approximate the solution of Poisson's equation in complex 2D geometries. These geometries are particularly relevant to the simulation of LTPs. The approach employs a hybrid CNN-transformer network architecture combined with a weighted multiterm loss function and a highly-randomized synthetic data generation scheme. The paper is structured as follows: in section~\ref{sec:scope}, we first briefly review recent works on solving Poisson's equation using ANNs, then we specify the problem definition and the scope of the present work. In section~\ref{sec:methods}, we describe the dataset preparation, objective function formulation, network architecture, and the details of training experiments. Subsequently, we present and discuss the results in section~\ref{sec:results_discussion}, which include the training success, model evaluations, and accuracy of the calculated electric fields from the predicted potential profiles as well as the wall time performance of the learned solvers. Finally, we conclude the work in section~\ref{sec:conclusion_outlook} and provide the outlook for future developments.

\section{Problem definition and scope}
\label{sec:scope}
Poisson's equation is often written in the following form:
\begin{equation}
\label{eq:poisson_eq}
    -\nabla^2 \phi \left(\boldsymbol{r}\right) = f\left(\boldsymbol{r}\right),
\end{equation}
where $f(\boldsymbol{r})$ is a known source term, $\boldsymbol{r}$ is the position vector, and $\phi(\boldsymbol{r})$ is the unknown variable to be solved. Boundary conditions such as Dirichlet, Neumann, and mixed types must be applied to solve the equation. Fundamentally, solving Poisson's equation with machine learning-based methods falls into the category of regression problems due to the continuous nature of the output. Typically for 2D cases, the prediction is done at once for all spatial points in the domain, and the input and output data are co-located and described in the same spatial configuration. To set the scope, we briefly review two recent developments in machine learning-based Poisson solvers for 2D cases using CNNs \cite{Ozbay-2021-ID6201,Cheng-2021-ID6203}. Both referenced works have set up the Poisson problem in this fashion. %\unsure{Furthermore, there are three important aspects to measure the quality of surrogate/machine-learned solvers: generalizability, accuracy, and time. In this work, we mainly focus on the first two aspects.}

Özbay \textit{et al.} \cite{Ozbay-2021-ID6201} have proposed a CNN-based model trained using a novel $L^p$-norm based loss function to solve the Poisson equation in fluid simulations of incompressible flows. The presented model was trained in a supervised learning scheme on a randomly generated dataset defined on Cartesian grids. The input data for the model consist of the source term, the domain size, and the boundary conditions (Neumann or Dirichlet boundary conditions). This expressive input data allows for a wide range of problem configurations that the trained model can solve. Additionally, they have demonstrated that a significant increase in accuracy can be achieved by using the predicted solution as an initial guess for a single iteration of a conventional iterative solver. The presented model, however, is not applicable to use cases with mixed boundary conditions and the presence of objects within the domain.

For electrostatic systems, Cheng \textit{et al.} \cite{Cheng-2021-ID6203} have investigated two well-established CNN architectures, and have studied the network hyperparameters well suited for solving Poisson's equation in fluid simulations of low-temperature plasmas. As a first step, the architectures have been trained to solve equation~\eref{eq:poisson_eq} with homogeneous Dirichlet boundary conditions and a given source term as input, in a supervised learning method similar to \cite{Ozbay-2021-ID6201}. The best-performing architecture, in this case the U-Net architecture \cite{Ronneberger-2015-ID6204}, has then been trained on simulation-specific datasets to solve the equation in \textit{real} settings such as plasma oscillation and double-headed streamer simulations. The trained model has been successfully evaluated on the corresponding simulation cases. Since the model uses only the source term as an input parameter, while boundary conditions and domain size are implied in the training data, this limits the range of problems (e.g., different reactor geometries) that the model can solve without retraining.

The learned Poisson solvers discussed above have so far dealt only with Poisson problems in simple geometries and in the absence of objects inside the computational domain. Furthermore, the boundary conditions are defined only on the outer boundaries of the domain and with limited configurations. This limits the usability of the learned solvers for application in LTP simulations without retraining the networks. For example in the simulation of a dielectric barrier discharge (DBD) or packed-bed reactors, one may incorporate inner electrodes and geometrically complex dielectric materials with varying coefficient values within the computational domain. Furthermore, mixed Neumann-Dirichlet boundary conditions are often encountered \cite{Jovanovic-2022-ID6205,Engeling-2018-ID6206,Zhang-2021-ID5808}. In addition, the electrodes may be dynamically driven, meaning that the boundary conditions (value or gradient) change over time \cite{Jovanovic-2022-ID6205,Baldry-2021-ID6207, Engeling-2018-ID6206,Zhang-2021-ID5808}.

In this work, we aim to investigate the feasibility of developing a 2D machine-learned Poisson solver that is \textit{generic} enough for most use cases in LTP simulations. Generic here means that the learned solver must be able to handle arbitrary charge density distributions, dielectric material distributions, domain sizes, boundary values, as well as different boundary conditions (Dirichlet, Neumann, and mixed) while avoiding the expensive cost of retraining. At this stage, we still restrict the problems to the use cases in LTP simulations with structured Cartesian grids and fixed numbers of computational nodes to make the task tractable. To account for the above-mentioned requirements, we consider Poisson's equation in 2D Cartesian geometries in the following form:
\begin{eqnarray}
\label{eq:generalized_poisson}
    -\nabla \cdot \left(\varepsilon_{\mathrm{r}}\left(x,y\right)\nabla \phi \left(x,y\right) \right) = \frac{\rho\left(x,y\right)}{\varepsilon_0},~\forall x,y \in \Omega \\
\label{eq:dirichlet_cond}
    \phi\left(x,y\right) = h\left(x,y\right),~\forall x,y \in \Gamma_\mathrm{Dirichlet} \\
\label{eq:neumann_cond}
    \frac{\partial}{\partial \boldsymbol{n}}\phi\left(x,y\right) = g\left(x,y\right),~ \forall x,y \in \Gamma_\mathrm{Neumann}
\end{eqnarray}
where $\varepsilon_{\mathrm{r}}$ denotes the spatial distribution of the dielectric constant, $\rho$ is the charge density, $\varepsilon_0$ is the vacuum permittivity, $\rho / \varepsilon_0 = f$ is the source term, $h$ is the Dirichlet boundary condition value defined on $\Gamma_\mathrm{Dirichlet} \subset \partial \Omega$, $g$ is the Neumann boundary condition value defined on $\Gamma_\mathrm{Neumann} \subset \partial \Omega$, and $\boldsymbol{n}$ is the unit normal to the boundary $\partial \Omega$ (both inner and outer boundaries). Lastly, $\Omega$ is the solution domain. The solution domain and the boundaries are defined on a rectangular domain with width $w$ and length $l$.

\begin{figure*}[tbp]\centering
\includegraphics[width=1\textwidth]{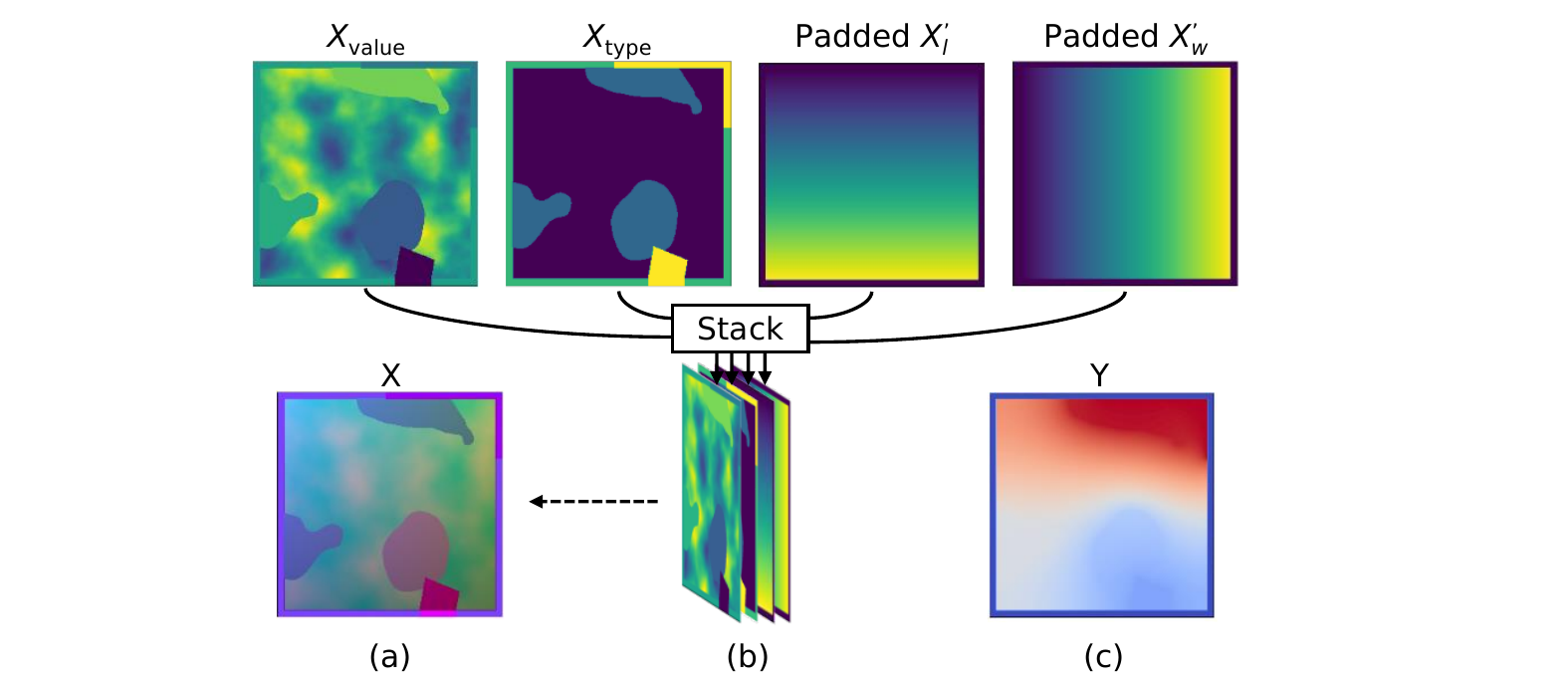} 
\caption{Input data configuration. (a) Configured input data presented as a CMYK image, (b) different components of input data, and (c) the corresponding output data (potential distribution).}
\label{fig:data_conf}
\end{figure*}

\section{Methods}
\label{sec:methods}
In connection with computer vision, discretized 2D Poisson problems defined in structured Cartesian grids can be thought of as pixel-wise regression problems, where input and output data are arranged in the spatial domain equivalent to images (or tensors). Consequently, each pixel in an image can be viewed as a computational node within the domain. Monocular depth estimation tasks are a good example, where a machine learning model must predict the depth value of each pixel in the output image from its input image. Like in \cite{Ozbay-2021-ID6201,Cheng-2021-ID6203}, we tackle this regression problem in a supervised learning manner, i.e., learning from input-output data pairs.

\subsection{Datasets}
\label{sec:dataset}
One of the expected features of the learned solver presented in this work is that it should generalize well on unseen data, i.e., data excluded from the training process. For the given task, this generalization capability is (among other aspects) indicated by how well the learned solver can handle reactor geometries of arbitrary configurations. It is well known that ANNs thrive on a large and high-quality dataset. However, generating such a dataset from actual numerical simulations using actual reactor geometries is a computationally expensive process and limits the diversity of the data. Therefore, we choose to use a random process to generate highly unique data as done similarly in \cite{Ozbay-2021-ID6201,Cheng-2021-ID6203,Li-2020-ID6192}. In this way, we can generate training data that are distinct from each other in terms of physical quantities and spatial features (shapes of objects within the domain) or geometries. This diverse training data encourage the network to learn the appropriate mapping between the input and the output spaces and prevents the network from over-memorization (overfitting). Thus, improving the overall generalization capabilities of the learned solver.

Let $\mathrm{X}\in\mathbb{R}^{H \times W \times C}$ be an input data sample and $\mathrm{Y}\in\mathbb{R}^{H \times W}$ be the corresponding output data sample, where $H$ is the number of pixels in the $y$ direction (height), $W$ is the number of pixels in the $x$ direction (width), and $C$ denotes the number of channels of the image. For example, a color image is often described in the RGB color space $\left(C=3\right)$, though other types of color space also exist such as CMYK $\left(C=4\right)$. The output data $\mathrm{Y}$ corresponds to the electrostatic potential which is a scalar field equivalent to a monochromatic image, e.g., a grayscale image. For solving equation \eref{eq:generalized_poisson}, the tensor $\mathrm{X}$ is composed of the different scalar input fields required to solve the equation such as $\rho/\varepsilon_0$, $\varepsilon_{\mathrm{r}}$, $g$, $h$, $\partial \Omega$ and $\Omega$ (see equations \eref{eq:generalized_poisson}--\eref{eq:neumann_cond}, detailed in section\,\ref{sssec:input_data}).

\subsubsection{Feature scaling}
% As per equations (\ref{eq:generalized_poisson}), (\ref{eq:dirichlet_cond}), and (\ref{eq:neumann_cond}), we consider a number of parameters $\varepsilon_{\mathrm{r}}$, $h$, $\rho$, $g$, and $\Omega =  [0,l] \times [0,w]$, which we explicitly define as the input parameters.
For real applications, the physical quantities of the mentioned parameters are of vastly different orders of magnitude. This may have severe effects on the learning success. Because ideally, every input parameter has to contribute equally in the gradient calculations during training \cite{Shanker-1996-ID6208,LeCun-1998-ID6209}. Therefore, we use the following feature scaling scheme. For the cases of LTP simulations, we can recast equation \eref{eq:generalized_poisson} into a dimensionless problem
\begin{equation}
\label{eq:normalized_poisson}
    - \tilde{\nabla} \cdot (\varepsilon_{\mathrm{r}}(\tilde{x}, \tilde{y}) \tilde{\nabla} \tilde{\phi}(\tilde{x}, \tilde{y})) = \nu \tilde{\rho}(\tilde{x}, \tilde{y}),
\end{equation}
\begin{equation}
\label{eq:normalization_coeff}
    \nu :=\frac{\breve{\rho}\breve{r}^2}{\varepsilon_0\breve{\phi}} = \mathrm{const\,},
\end{equation}
where $\nu$ is a scaling coefficient, while quantities denoted by $\tilde{*}$ and $\breve{*}$ are the scaled quantity and the scaling factor of the corresponding variable, respectively. The \mbox{$\nabla$-operator} is scaled using a typical length scale $\breve{r}=(\breve{x}\breve{y})^{1/2}$ as scaling factor. The scaling factors may be obtained from representative LTP discharge parameters, with the goal to obtain quantities of the order of unity. The scaling coefficient $\nu$ may be estimated correspondingly.

Consider the example of a two-dimensional low-pressure capacitively coupled radio-frequency (CCRF) discharge with an approximate maximum charge density in the sheath regions of $\breve{\rho} \approx 5 \cdot 10^{-5} \, \textrm{C}/\textrm{m}^3$, a powered electrode voltage of $\breve{\phi} \approx 250 \, \textrm{V}$, and a plasma reactor of size $\breve{x} \times \breve{y} = w \times l = 25 \times 50 \, \textrm{mm}^2$. For the sheath region, it may be estimated that $\nu \approx 56$. In contrast, the quasi-neutral LTP bulk typically extends over a large part of the domain. To also accommodate for $\breve{\rho}\ll 5 \cdot 10^{-5} \, \textrm{C}/\textrm{m}^3$ in this region, a global scaling factor of $\nu=1$ is chosen. While for the sheath regions, this corresponds to typical values for the scaled potential of $\tilde{\phi} \approx 56$, typical values of $\tilde{\phi} \lesssim 0.1$ are expected in the bulk. Note that the remaining scaled quantities $\tilde{*}$ are within the range $[-1, 1]$, which we will use for generating the training data.

\begin{figure*}[tbp]\centering
\includegraphics[width=1\textwidth]{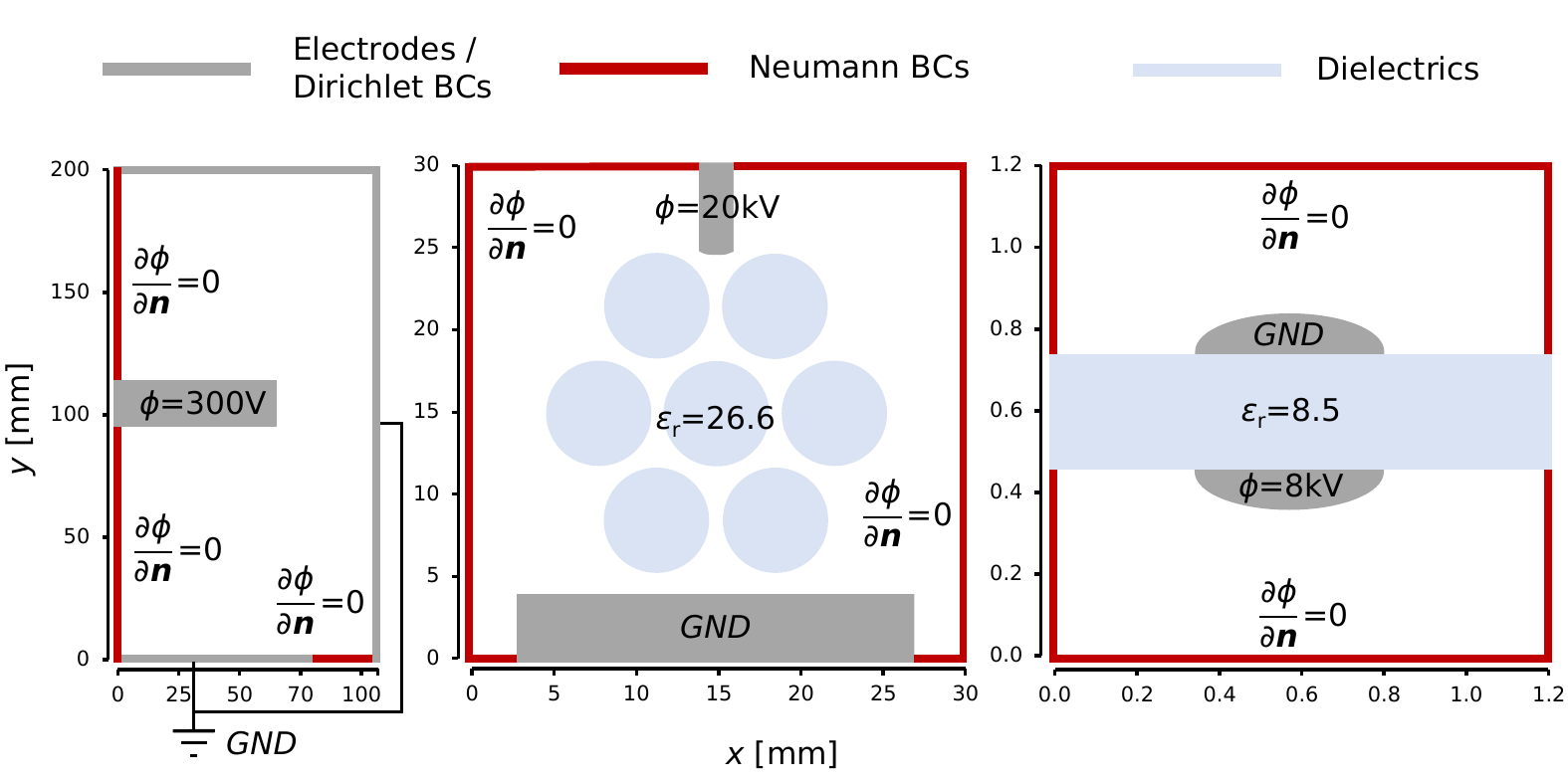} 
\caption{Reference LTP reactor geometries from literature. From left to right: asymmetric CCRF discharge \cite{Baldry-2021-ID6207}, packed-bed DBD \cite{Engeling-2018-ID6206}, and surface DBD \cite{Zhang-2021-ID5808} reactor geometries, which are used to generate $D_{*}^{(\mathrm{ccrf})}$, $D_{*}^{(\mathrm{dbd1})}$, and $D_{*}^{(\mathrm{dbd2})}$, respectively. BCs stands for boundary conditions.}
\label{fig:known_geo}
\end{figure*}

We can now define all input parameters $\tilde h$, $\tilde g$, $\tilde \rho$, $\tilde w$, and $\tilde l$ without physical dimensions and within the same range (except for $\varepsilon_{\mathrm{r}}$ which is always dimensionless). Here, we define the lower and upper bounds of the input parameters, $\tilde h$, $\tilde g$, $\tilde \rho \in [-1,1]$, and $\tilde w$, $\tilde l \in [0,1]$. In the case of $\varepsilon_{\mathrm{r}}$, we choose $\varepsilon_{\mathrm{r}} \in [2,30]$. This is because, firstly, the dielectric materials must have $\varepsilon_{\mathrm{r}} > 1$, since $\varepsilon_{\mathrm{r}} = 1$ is the dielectric constant for free space. Secondly, we consider the practical use cases in LTP, which rarely use dielectric materials that have $\varepsilon_{\mathrm{r}} > 30$. Note that the described scaling specification is chosen for the sake of simplicity and may not cover every use case in LTP simulations. However, for practical and specific use cases, the specification can be adjusted accordingly with ease, i.e., by adjusting $\nu$ and the upper and lower bounds of the input parameters.

\subsubsection{Data generation and normalization}
Objects inside the computational domain are made up of $\varepsilon_{\mathrm{r}}$ or $\tilde h$ or a combination thereof. For these two parameters, we generate instances that have random shapes and values within the defined range. Additionally, $\tilde h$ can also be defined at the outer domain boundary, and in this case, we randomly define the nodes (equivalent to pixels in an image) at the outer boundary with values of $\tilde h$. This also applies to $\tilde g$, which is exclusively defined at the outer boundary. In the case of $\tilde \rho$, unfortunately, it is not a straightforward task to procedurally generate \textit{fake} charge density distributions that are physically plausible. To this end, we use 2D random noise generation methods (such as Perlin and Simplex noises)\cite{githubGitHubRobbmcleodpyfastnoisesimd}, considering that it may capture the features of \textit{real} physical distributions. Lastly, we randomly define the spatial dimension $\tilde w$ and $\tilde l$. We generate the input parameters as images of the same size, $X_{\varepsilon_{\mathrm{r}}}$, $X_\rho$, $X_h$, $X_g$, $X_w$, $X_l \in \mathbb{R}^{U\times V}$, where $U \times V$ is the size of the images. Here, relevant information in $X_{\varepsilon_{\mathrm{r}}}$, $X_\rho$, $X_h$, and $X_g$ do not overlap with each other. For example, we consider that there are no electric charges within dielectric materials and that dielectric and electrode materials are clearly distinct. Therefore, we set the values of all non-relevant pixels in the images to zero. As for $X_w$ and $X_l$, we discretize $\tilde \Omega \cup \partial \tilde \Omega = [0,\tilde  w] \times [0,\tilde l]$ such that $x_w^{mn} = m \Delta x_w$ and $x_l^{mn} = n \Delta x_l$, where $ \Delta x_w = \tilde w /(V-1)$, $\Delta x_l = \tilde l /(U-1)$, $x_{w}^{mn}$ is the pixel value at an $(m,n)$ coordinate point in $X_w$, and similarly for $x_l^{mn}$. Furthermore, each element in $X_{\varepsilon_{\mathrm{r}}}$, $X_\rho$, $X_h$, $X_g$, $X_w$, and $X_l$ is co-located with each other meaning that every element with the same coordinates directly contributes to the prediction outcome that shares the same coordinates. Using the generated input data, we can create the corresponding solution data $\tilde \phi$ by solving it using a conventional PDE solution method. In this case, we use a second-order finite difference method (FDM) solver implemented in the NumPy (version 1.24) and SciPy (version 1.9.3) libraries \cite{Harris-2020-ID6210,Virtanen-2020-ID6211, githubGitHubNumpynumpy, githubGitHubScipyscipy}. Similarly, the solution data is arranged as an image $Y \in \mathbb{R}^{U\times V}$.

To further ease the learning process, we apply the min-max normalization to all input data effectively scaling their values to a uniform range of $[0, 1]$. As for the solution data, the min-max normalization is done by using the minimum and maximum values from the corresponding $\tilde h$, since $\tilde \phi$ is unbounded. In practice, we found that almost all normalized solution data are within $[0,1]$. Thus, we obtain the normalized data (i.e., images) $X'_{\varepsilon_{\mathrm{r}}}$, $X'_\rho$, $X'_h$, $X'_g$, $X'_w$, $X'_l \in [0,1]^{U\times V}$ and $Y' \in \sim [0,1]^{U\times V}$. This final normalization reduces the size of the search space to mostly positive values further increasing learning efficiency.

%\textbf{Data generation and normalization.} As per (\ref{eq:generalized_poisson}), (\ref{eq:dirichlet_cond}), and (\ref{eq:neumann_cond}), we have a number of primary input parameters $\varepsilon_{\mathrm{r}}$, $h$, $f$, $g$, and $\Omega$, which are all described spatially. Objects inside the computational domain $\Omega$ are made up of $\varepsilon_{\mathrm{r}}$ or $h$ or a combination thereof. For these two parameters, we generate instances that have random shapes and values. Additionally, $h$ can also be defined at the domain boundary $\partial \Omega$, and in this case, we randomly define the nodes/pixels at $\partial \Omega$ with values of $h$. This also applies to $g$, which is exclusively defined at $\partial \Omega$. In the case of $f$, we use 2D random noise generation methods (such as Perlin and Simplex noise \textbf{[cite,cite]}) with the hope that it may \unsure{(\textit{approximately})} capture the \textit{real} physical distributions of $f$. Here, we define $\varepsilon_0 = 1$, effectively making $f=\rho$. Finally, we randomly define the physical dimension of the domain $\Omega = [0,l] \times [0,w]$, where $l$ is the length and $w$ is the width of the domain, the values that are randomized. So far, the input data generation is described. And by using the generated input data, we can create the corresponding output data by solving it using a conventional method. In this case, we use a direct second-order accurate finite difference method.

\subsubsection{Input data configuration}
\label{sssec:input_data}
We can exploit the co-location and non-overlapping properties of the input parameters to configure the input data efficiently. First, we apply edge-padding operations with pad size $p$ to $X'_h$, $X'_g$, $X'_\rho$, and $X'_{\varepsilon_{\mathrm{r}}}$. This is mainly done to increase the contributions of $\tilde g$ and $\tilde h$ defined at the outer boundary. Because in the images, this is equivalent to lines at the edges with a thickness of one pixel, which is relatively small. For $X'_w$ and $X'_l$, we apply zero-padding operations with the same $p$. As mentioned previously, the relevant information in $X'_h$, $X'_g$, $X'_\rho$, and $X'_{\varepsilon_{\mathrm{r}}}$ do not overlap with each other. This allows us to arrange them together into one image without losing any important information, and by doing so, we obtain the \textit{value} image $X_\mathrm{value} \in [0, 1]^{(U+2p)\times (V+2p)}$. Up to this point, it is unlikely for the network to be able to identify the different types of parameters in $X_\mathrm{value}$. To this end, we construct a \textit{type} image $X_\mathrm{type} \in [0, 1]^{(U+2p)\times (V+2p)}$, where each type of input parameter is assigned a distinct value in [0,1]. And in this case, we choose $\mathrm{type}_\rho = 0.25$, $\mathrm{type}_{\varepsilon_{\mathrm{r}}} = 0.5$, $\mathrm{type}_g = 0.75$ and $\mathrm{type}_h = 1.0$. Finally, we can stack $X_\mathrm{value}$,  $X_\mathrm{type}$, and the padded $X'_l$ and $X'_w$ together into one final input data $\mathrm{X} \in [0, 1]^{(U+2p)\times (V+2p) \times 4}$. In fact, $U + 2p = H$ and $V + 2p = W$ is the final size of the input data, thus, $\mathrm{X} \in [0, 1]^{H\times W \times C}$ with $C = 4$. Figure \ref{fig:data_conf} depicts the general overview of the input data configuration. Figure \ref{fig:data_conf}(a) shows an example of input data visualized in the CMYK color space, and figure \ref{fig:data_conf}(b) shows the different components of the input data and how it can be stacked together to create the final input data. Lastly, to keep the uniformity throughout the whole process, we apply zero-padding with the same $p$ to $Y'$, which results in the final output data $\mathrm{Y} \in \sim [0,1]^{H\times W}$. An example of an output image is shown in figure \ref{fig:data_conf}(c).

\subsubsection{Prepared datasets}
For a complete dataset, we generate a number of subsets for training, validation, and testing or evaluation of the learned solvers. For training and validation, we generate two main sets $\mathcal{D}_{272}$ and $\mathcal{D}_{528}$. Here, $\mathcal{D}_{272}$ contains $N = 5\cdot 10^4$ input-output data pairs with $H \times W=272^2$ ($U\times V=256^2$ and $p=8$), and $\mathcal{D}_{528}$ contains $N = 3\cdot 10^4$ input-output data pairs with $H\times W=528^2$ ($U \times V=528^2$ and $p=8$). For testing or evaluation, we generate two sets, $\mathcal{D}_{272}^\mathrm{(test)}$ and $\mathcal{D}_{528}^\mathrm{(test)}$, each with $N=5\cdot 10^3$.

Additionally, we prepare six datasets from three reference reactor geometries loosely based on \cite{Baldry-2021-ID6207,Engeling-2018-ID6206,Zhang-2021-ID5808}, $\mathcal{D}_{272}^{(\mathrm{ccrf})}$, $\mathcal{D}_{528}^{(\mathrm{ccrf})}$, $\mathcal{D}_{272}^{(\mathrm{dbd}1)}$, $\mathcal{D}_{528}^{(\mathrm{dbd}1)}$, $\mathcal{D}_{272}^{(\mathrm{dbd}2)}$, and $\mathcal{D}_{528}^{(\mathrm{dbd}2)}$ for further evaluation of the learned solvers. Each of these sets has $N= 10^3$. Note that the reference geometry datasets are not obtained from real numerical simulations. Random processes are still used to generate the charge density and most of the input parameter values following the same scheme as the random datasets. Figure \ref{fig:known_geo} presents the known reactor geometries and their configurations.

%random processes are still used to generate the charge density and the rest of the input parameter values following the same scheme as the random datasets.

\subsection{Objective function formulation}
\label{sec:objective_function}
Now, we assume that there is a function $G$ that can approximate the mapping between $\mathrm{X}$ and $\mathrm{Y}$ to a certain degree of accuracy. The function $G$ is to be learned using an ANN. $G$ learns from a dataset containing $N$ observations of input-output pairs $\{\mathrm{X}_i,\mathrm{Y}_i\}_{i=1}^N$ obtained from solving $\mathrm{X}$ using a conventional method such as FDM. We therefore can formulate the objective function of the training procedure as an optimization problem that minimizes a loss function $\mathcal{L}$
\begin{equation}
\label{eq:objective_func}
    \min_\theta \mathbb{E}_{(\mathrm{X},\mathrm{Y})}\left[\mathcal{L}\left(\mathrm{Y},G_\theta\left(\mathrm{X}\right)\right)\right],
\end{equation}
where $\theta$ denotes trainable parameters of the neural network $G_\theta$ (a function $G$ parameterized by $\theta$). For a finite number of $N$ observations of training data, equation \eref{eq:objective_func} can be written as
\begin{eqnarray}
    \min_\theta \mathbb{E}_{(\mathrm{X},\mathrm{Y})}\left[\mathcal{L}\left(\mathrm{Y},G_\theta\left(\mathrm{X}\right)\right)\right] \nonumber \approx \min_\theta \frac{1}{N} \sum_{i=1}^{N} \mathcal{L}\left(\mathrm{Y}_i,G_\theta\left(\mathrm{X}_i\right)\right),
\end{eqnarray}
where $\mathrm{Y}_i$ and $\mathrm{X}_i$ are the $i$-th true output and $i$-th input data, respectively. Finally, $G_\theta\left(\mathrm{X}_i\right):=\hat{\mathrm{Y}}_i$ denotes the $i$-th predicted output.

For successful learning, the loss function must be designed appropriately. The most ubiquitous loss function for regression problems is the $L^{1}$-norm loss, which directly measures the distance between the true and predicted values at each pixel from the absolute differences of their coordinates, defined by
\begin{equation}
\label{eq:l_mae}
    \mathcal{L}_{L^1}(\mathrm{Y},\hat{\mathrm{Y}}) = \| \mathrm{Y} - \hat{\mathrm{Y}} \|_{1}.
\end{equation}
One property that the predicted output (solution to the Poisson equation) must have is \textit{smoothness}, both in terms of mathematical and perceived smoothness. Inspired by several works in monocular depth estimation tasks, we consider a single-scale structural similarity index measure (SSIM) \cite{Wang-2004-ID6212} loss
\begin{equation}
\label{eq:l_ssim}
    \mathcal{L}_\mathrm{SSIM} (\mathrm{Y},\hat{\mathrm{Y}})= 1 - \textnormal{SSIM}(\mathrm{Y},\hat{\mathrm{Y}})
\end{equation}
combined with the disparity smoothness loss \cite{Godard-2016-ID6213, Godard-2018-ID6214}
\begin{eqnarray}
\label{eq:l_smooth}
    \mathcal{L}_\mathrm{smooth}(\mathrm{Y},\hat{\mathrm{Y}}) = \|\partial_x \hat{\mathrm{Y}}\|_{1} \exp(-\|\partial_x\mathrm{Y}\|_{1}) & + \|\partial_y \hat{\mathrm{Y}}\|_{1} \exp(-\|\partial_y\mathrm{Y}\|_{1}).
\end{eqnarray}
Contributions of equations \eref{eq:l_ssim} and \eref{eq:l_smooth} have been shown to enforce the network to produce predictions that are perceptually similar to the true output and with correct smoothness \cite{Godard-2016-ID6213, Godard-2018-ID6214}. To the best of our knowledge, the use of $\mathcal{L}_\mathrm{SSIM}$ and $\mathcal{L}_\mathrm{smooth}$ have not yet been reported for PDE-related problems. Referring to the parameters of SSIM described in \ref{appendix:ssim}, we set the Gaussian window size to $3 \times 3$ and the dynamic range value $\lambda$ to 1.0 for equation \eref{eq:l_ssim}. Note that $\lambda$ is defined as the difference between the maximum and the minimum of possible output values. In this work, we found that the output values reside mostly within $[0,1]$ (see section~\ref{sec:dataset}), which is why we opt for using $\lambda = 1.0$.

Furthermore, we consider a contextual physics loss, which in this case is defined by the relative $L^2$-norm loss between the gradients of the true and predicted potentials
\begin{equation}
\label{eq:l_efield}
    \mathcal{L}_{\boldsymbol{E}} (\mathrm{Y},\hat{\mathrm{Y}}) = \frac{\|\nabla (\hat{\mathrm{Y}} - \mathrm{Y} ) \|_2^2}{\|\nabla \mathrm{Y}\|_2^2}.
\end{equation}
It measures the accuracy of the predicted electric field from the predicted potential $\hat{\mathrm{Y}}$, which is indeed the ultimate goal for the field calculation step in self-consistent electrostatic LTP simulations. It has been shown in \cite{Cao-2021-ID6193,Wen-2022-ID6170} that a relative loss term such as equation \eref{eq:l_efield} has a positive regularization effect on the training.

Finally, we can write the total loss function as a sum of these loss terms
\begin{equation}
\label{eq:l_total}
\mathcal{L}(\mathrm{Y},\hat{\mathrm{Y}}) = \alpha \mathcal{L}_{L^1}(\mathrm{Y},\hat{\mathrm{Y}}) + \beta \mathcal{L}_\mathrm{SSIM}(\mathrm{Y},\hat{\mathrm{Y}}) \nonumber + \gamma \mathcal{L}_\mathrm{smooth}(\mathrm{Y},\hat{\mathrm{Y}}) + \delta \mathcal{L}_{\boldsymbol{E}}(\mathrm{Y},\hat{\mathrm{Y}}),
\end{equation}
where $\alpha$, $\beta$, $\gamma$, and $\delta$ are the weights or hyperparameters. This weighted multiterm loss function is designed to enforce the network to produce both numerically accurate and smooth predictions. We found that setting $\alpha$, $\beta$, $\gamma$, and $\delta$ to $0.8$, $1.0$, $0.9$, and $0.8$, respectively, gives satisfactory results.

\subsection{Network architecture}
%CNNs have been used predominantly for dealing with data that can be represented as images or tensors mainly for computer vision tasks. As of late, the transformer networks, originally designed for machine-translation tasks \textbf{[Original Trans. paper]}, have been seeing strong adoption in computer vision prompted by the work of \textbf{[ViT paper]}. Transformer-based vision models are rapidly replacing state-of-the-art CNN-based models in many computer vision tasks \textbf{[cite all vision sotas]}, this is mainly due to their ability to strongly encode global contexts \unsure{(of the images)} and a high modeling capacity of large datasets \textbf{[cite,cite]}. For this task, we employ a slightly modified version of a hybrid CNN-transformer architecture from \textbf{[TransUnet]} dubbed TransUNet, which was initially designed for medical image segmentation tasks.
CNNs have been used predominantly for dealing with data that can be represented as images or tensors mainly for computer vision tasks. Recently, transformer networks, originally designed for machine-translation tasks by Vaswani \textit{et al.} \cite{Vaswani-2017-ID6181}, have been widely adopted in computer vision prompted by the work of Dosovitskiy \textit{et al.} \cite{Dosovitskiy-2020-ID6215}, and rapidly replacing state-of-the-art CNN-based models in many computer vision tasks \cite{Yu-2022-ID6216, Chen-2022-ID6217, Chen-2022-ID6218, Xie-2022-ID6219}. In this work, we employ a slightly modified version of a hybrid CNN-transformer architecture proposed by Chen \textit{et al.} \cite{Chen-2021-ID6220} dubbed TransUNet, which was initially designed for medical image segmentation tasks.

\begin{figure*}[tbp]\centering
\includegraphics[width=1.0\textwidth]{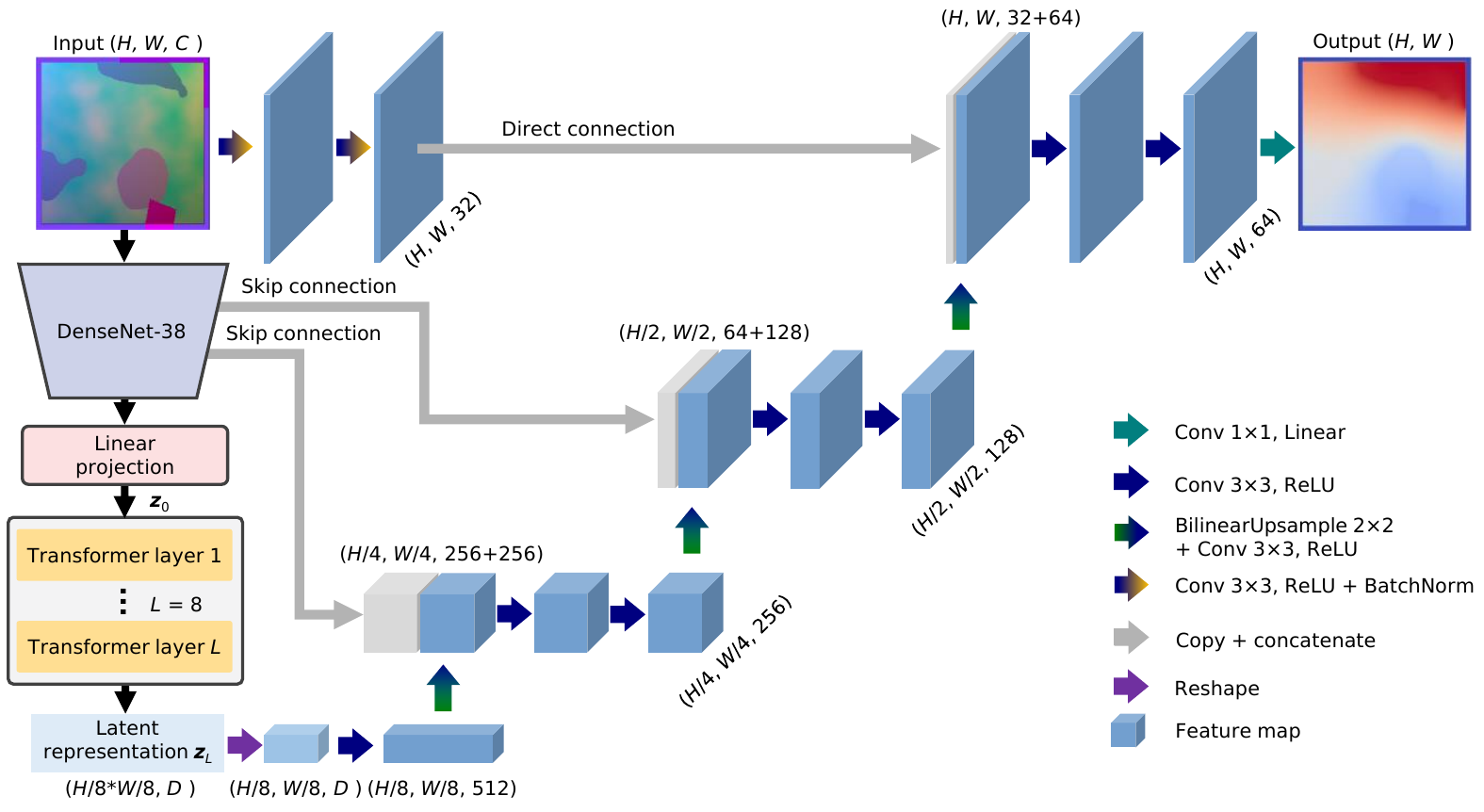} 
\caption{Overview of the TransDenseUNet architecture. The encoder incorporates DenseNet and transformer networks, whereas the decoder is made of the typical CNN blocks.}
\label{fig:architecture}
\end{figure*}

TransUNet follows the u-shaped design with skip connections introduced by Ronneberger \textit{et al.} in \cite{Ronneberger-2015-ID6204}. The design consists of encoder and decoder parts, where feature information from encoding stages gets relayed directly to the corresponding decoding stages via skip connections. In TransUNet, the transformer layers are exclusively used in the encoder part in conjunction with the typical convolution and pooling layers, while the decoder part follows the usual CNN-based decoder design. The hybridization of transformers and CNNs supported by the skip connections leads to the ability of TransUNet to encode strong global and low-level features \cite{Chen-2021-ID6220}. This is one of the primary reasons why we presume such an architecture design is well suited for the given problem, considering its elliptic nature. In the following, we describe in more detail the modified TransUNet architecture used in this work, which we call TransDenseUNet (TDN). The high-level overview of the architecture is depicted in figure \ref{fig:architecture}.

\subsubsection{Encoder} 
\begin{figure}[tbp]\centering
\includegraphics[width=0.5\columnwidth]{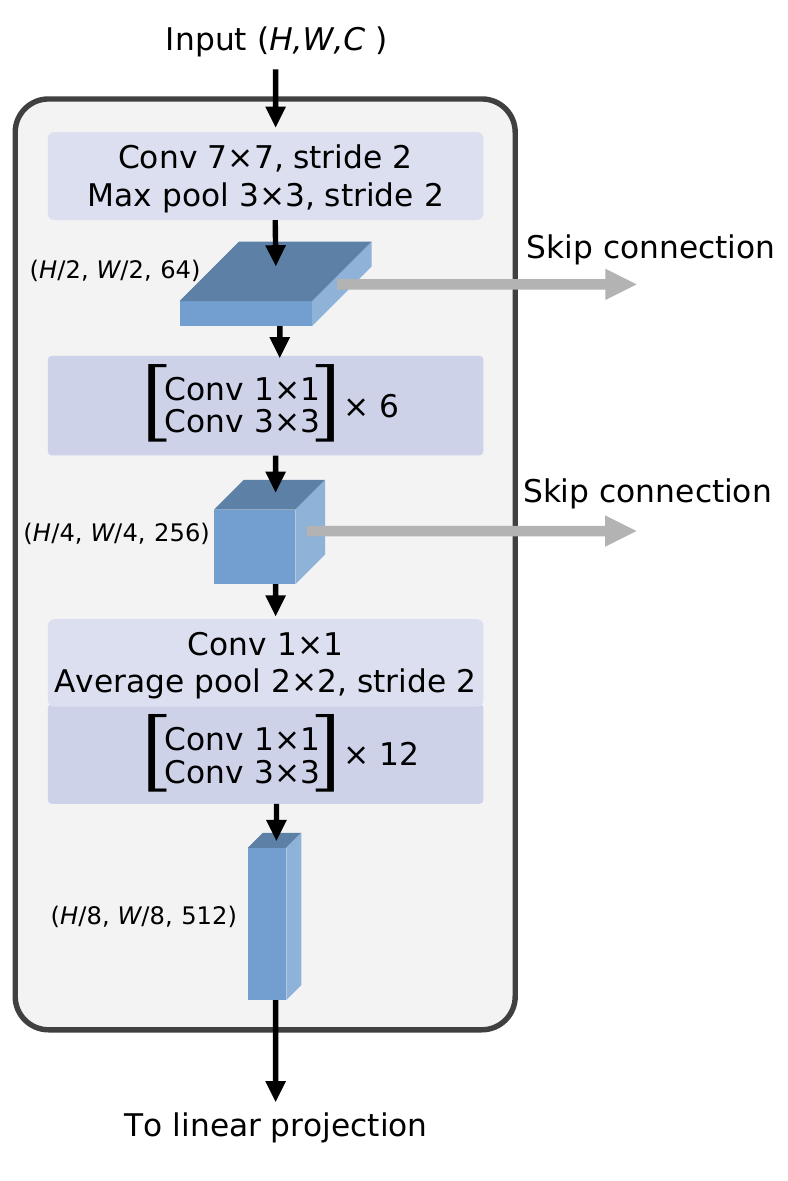} 
\caption{High-level structure of the DenseNet-38 block used in the TransDenseUNet. It consists of 38 densely connected convolutional layers.}
\label{fig:densenet38}
\end{figure}

The input data of size $H \times W \times C$ is first downsampled through several convolution and pooling operations. Unlike in the original implementation \cite{Chen-2021-ID6220}, we employ a block of densely connected convolutional layers following the DenseNet architecture \cite{Huang-2017-ID6221} to downsample the input data in place of regular CNN layers. The DenseNet approach offers several advantages such as enabling training deeper layers with a reduced number of parameters and strengthening feature propagation, which enables efficient and powerful representation learning \cite{Huang-2017-ID6221}. In total, we employ 38 convolutional layers that are densely connected, thus, we call this first encoder part the DenseNet-38 block. Intermediate outputs are employed as skip connections to the decoder network. The final output of this block is a feature map of size $\frac{H}{8}\times \frac{W}{8} \times 512$. A high-level structure of DenseNet-38 is presented in Figure \ref{fig:densenet38}. For further details regarding the DenseNet architecture, we refer the reader to \cite{Huang-2017-ID6221}.

\begin{figure}[tbp]\centering
\includegraphics[width=0.5\columnwidth]{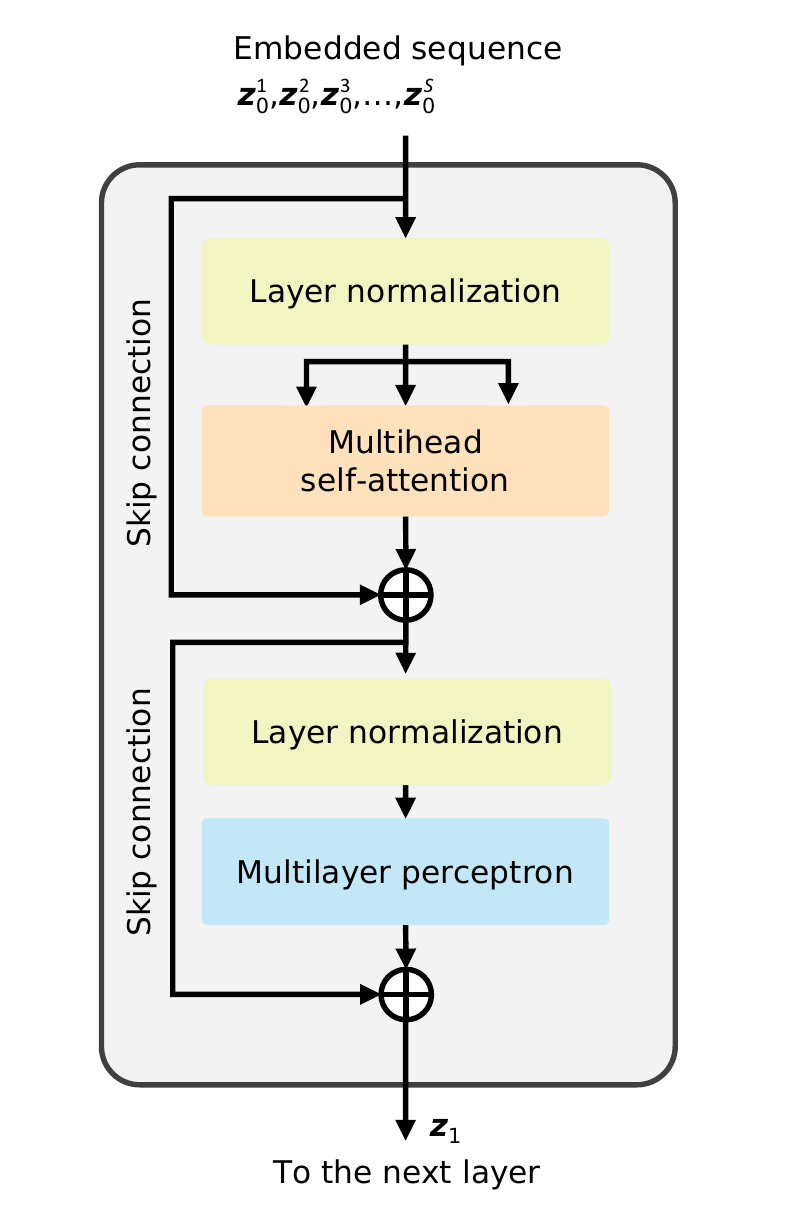} 
\caption{General structure of a transformer layer used in the TransDenseUNet architecture.}
\label{fig:transformer_layer}
\end{figure}

From here on, the encoding process is carried out by the transformers. A standard transformer layer requires a 1D input sequence, therefore, the outputted 2D features must be transformed into a sequence of flattened patches $\boldsymbol{x}_p \in \mathbb{R}^{S \times 512}$, where $S=\frac{HW}{8^2P^2}$ is the resulting number of patches, and $P \times P$ is the dimension of each patch. Here, we opt for $P=1$ resulting in $S=\frac{HW}{8^2}$, and at this stage, we obtain the tokenized feature maps (or patches) $\boldsymbol{x}_p$. Following \cite{Chen-2021-ID6220,Dosovitskiy-2020-ID6215}, we evaluate patch and position embeddings using a trainable linear projection, which transforms $\boldsymbol{x}_p$ into an embedded sequence $\boldsymbol{z}_0 \in \mathbb{R}^{S \times D}$, where $D$ is the embedding dimension. Finally, the embedded sequence $\boldsymbol{z}_0$ is the input to the transformer block, which contains $L$ transformer layers. Each transformer layer consists of a multihead self-attention (MSA) block with $N_\mathrm{MSA}$ self-attention heads, a multilayer perceptron (MLP) block, two-layer normalization operations, two skip connections, and two vector addition operations, structured as shown in Figure \ref{fig:transformer_layer}. The MLP block has two layers of ANNs, where the first layer has $N_\mathrm{neuron}$ neurons and the last layer has the same number of neurons as the embedding dimension $D$. Finally, from the last transformer layer, we obtain the final latent representation $\boldsymbol{z}_L \in \mathbb{R}^{S \times D}$, which is ready to be reconstructed into a prediction by the decoder. For this task, $D=256$, $L=8$, $N_\mathrm{MSA}=4$, and $N_\mathrm{neuron} = 1024$ provide a good balance between prediction accuracy and computational cost (as well as memory requirement).  We refer interested readers to \cite{Vaswani-2017-ID6181, Dosovitskiy-2020-ID6215} for more in-depth descriptions of the transformer architecture.

The self-attention mechanism is often regarded as the key factor in the transformer architecture. As described in \cite{Vaswani-2017-ID6181}, MSA enables the network to pay attention to different pieces of information in the input sequence and utilize them efficiently. In natural language processing (NLP), MSA may consider different parts of the input text (a text corpus, a page in a book, etc.) and learn the global context. In \cite{Cao-2021-ID6193}, Shuhao Cao has presented a modified version of the self-attention mechanism and successfully applied it in an operator learning task to solve an elliptic BVP such as the Darcy flow problem.

\subsubsection{Skip connections} Skip connections play an important role since a significant number of the training data contains objects inside the domain. For such data the input and the reciprocal output share similar spatial features like shapes of the inner electrodes and dielectric materials, which can occupy a considerable space in the domain. The use of skip connections ensures the preservation of these spatial features from high to low spatial resolutions, connecting different stages of the downsampling process to the corresponding upsampling process (in each of the expansion stages) \cite{Ronneberger-2015-ID6204, Chen-2021-ID6220}.

For the presented architecture, the downsampling process is primarily carried out by the DenseNet-38 block \cite{Huang-2017-ID6221}. From this process, we extract two feature maps of sizes $\frac{H}{2}\times\frac{W}{2}\times 64$ and $\frac{H}{4}\times\frac{W}{4}\times 256$ (referred to as skip connections in Figure \ref{fig:densenet38}) and concatenate them to the corresponding upsampled feature maps that have the same spatial sizes. Additionally, we apply (two times) a one-stride $3 \times 3$ convolution with 32 filters to the input data, followed by a ReLU activation \cite{Fukushima-1975-ID6222, Nair-2010-ID6225} and a batch normalization layer \cite{Ioffe-2015-ID6226}. This results in a feature map of size $H\times W \times 32$, which we use for the concatenation of features that have the original spatial size $H\times W$. Note that this additional process is not connected to the down-/upsampling process, and is solely done for superposition of high-resolution features. In total, this architecture has two levels of skip connections and concatenation processes, plus one direct connection and concatenation process.

%The downsampling process is typically done in $k$-stages (depending on the size of the input image), and in such a way that a $k$-th stage results in a feature map of size $\frac{H}{2^k} \times \frac{W}{2^k} \times C_k$ achieved by a pooling operation, where $C_k$ is the number of the $k$-th feature map channel. And for each downsampling stage, the output feature map is copied and then concatenated to the corresponding upsampling stage that has the same spatial feature map size. Here, the downsampling process is primarily done by the dense blocks.

\subsubsection{Decoder} Ideally, the latent representation $\boldsymbol{z}_L$ obtained from the encoder contains all relevant information about the prediction. It is the task of the decoder to reconstruct $\boldsymbol{z}_L\in \mathbb{R}^{S\times D}$ into the prediction $\hat{\mathrm{Y}}\in \mathbb{R}^{H \times W}$. The decoder consists of several expansion stages. At the first step, we reshape $\boldsymbol{z}_L$ into a 2D feature map of size $\frac{H}{8} \times \frac{W}{8} \times D$, similar to the original implementation \cite{Chen-2021-ID6220}. After that, we run the feature map through three expansion stages, where the $k$-th expansion stage results in a feature map of size $\frac{H}{2^{3-k}} \times \frac{W}{2^{3-k}} \times \frac{512}{2^{k}}$. For each expansion stage, we apply a set of operations successively very similar to the implementation in \cite{Ronneberger-2015-ID6204}. Here, we opt for using a $2 \times 2$ bilinear-upsampling operation followed by a one-stride $3 \times 3$ convolution operation and a ReLU activation in the expansion operation (green-blue arrows in figure \ref{fig:architecture}). The concatenation of features takes place directly after the expansion operation, which is then followed by two times of a one-stride $3\times3$ convolution operation and a ReLU activation. Finally at the end of the last expansion stage, we apply a one-stride $1 \times 1$ convolution operation followed by a linear activation. This operation concludes the decoding process, which transforms a feature map of size $H \times W \times 64$ to the reconstructed output prediction $\hat{\mathrm{Y}} \in \mathbb{R}^{H \times W}$.

\subsection{Training experiment details}
In this work, we train several models primarily to assess the influence of the number and the resolution of training data on the model performance. To this end, we prepare sub-datasets $\mathcal{D}_{272}^{(\mathrm{10k})},\mathcal{D}_{272}^{(\mathrm{30k})} \subset \mathcal{D}_{272}$, $\mathcal{D}_{272}^{(\mathrm{50k})}= \mathcal{D}_{272}$, and use $\mathcal{D}_{528}$ as it is. In total, we obtain four sub- and datasets for training and validation. We split each set into a training set (80\% of the total data) and a validation set (20\% of the total data). Four models are trained each using one of the prepared datasets, and then evaluated on the prepared evaluation sets (as described in section~\ref{sec:dataset}) according to its data resolution. The models and the corresponding datasets as well as the number of trainable parameters are summarized in table \ref{table:models}. TDN528$_\mathrm{30k}$ has slightly more trainable parameters than the rest of the models due to the increase in resolution.

\begin{table*}
\centering \small
\caption{\label{table:models}TDN models and the corresponding datasets for training and validation as well as numbers of trainable parameters.}
\begin{tabular*}{1\columnwidth}{@{}l*{15}{@{\extracolsep{0pt plus12pt}}l}}
\br
         & & \centre{1}{$\#$ train.} & \centre{1}{\# val.}                 & \centre{1}{\# trainable}\\
Model    &\centre{1}{Dataset}     & \centre{1}{data}   & \centre{1}{data} & \centre{1}{parameters}  \\
\mr
TDN272$_\mathrm{10k}$ & \centre{1}{$\mathcal{D}_{272}^{(\mathrm{10k})}$} & \centre{1}{8k}  & \centre{1}{2k}  & \centre{1}{18.54m}\\
TDN272$_\mathrm{30k}$ & \centre{1}{$\mathcal{D}_{272}^{(\mathrm{30k})}$} & \centre{1}{24k} & \centre{1}{6k}  & \centre{1}{18.54m}\\
TDN272$_\mathrm{50k}$ & \centre{1}{$\mathcal{D}_{272}^{(\mathrm{50k})}$} & \centre{1}{40k} & \centre{1}{10k} & \centre{1}{18.54m}\\
TDN528$_\mathrm{30k}$ & \centre{1}{$\mathcal{D}_{528}$}         & \centre{1}{24k} & \centre{1}{6k}  & \centre{1}{19.359m}\\
\br
\end{tabular*}
\end{table*}

\begin{figure}[tbp]\centering
\includegraphics[width=0.5\columnwidth]{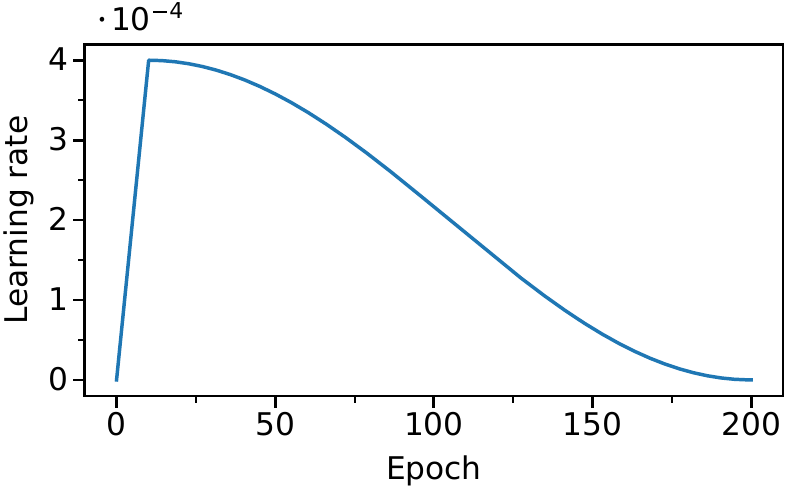} 
\caption{Cosine decay learning rate schedule with warm-up steps. Hyperparameters: $lr_\mathrm{start}= 0.0$, $lr_\mathrm{max}= 4\cdot 10^{-4}$, epoch $= 200$, batch size $= 8$, hold $= 10$, and warm-up steps $= 10 \cdot s$.}
\label{fig:cosind_schedule}
\end{figure}

For all training experiments, we choose the Adam optimizer \cite{Kingma-2014-ID6227} and the cosine decay learning rate schedule \cite{Loshchilov-2016-ID6223} with a warm-up. For this learning rate schedule, we set the hyperparameters as follows: start learning rate $lr_\mathrm{start}= 0.0$, maximum learning rate $lr_\mathrm{max}= 4\cdot 10^{-4}$, epoch $= 200$, batch size $= 8$, hold $= 10$, and warm-up steps $= 10 \cdot s$, where $s$ is the number of steps per epoch. The learning rate schedule is shown in figure \ref{fig:cosind_schedule}. During training, we apply a basic (image-)data augmentation process consisting of random flipping and rotation operations. Finally, we use the relative $L^2$-norm error $\varepsilon_{L^2}$ as an evaluation metric during and after the training, given by
\begin{equation}
\label{eq:rel_error}
    \varepsilon_{L^2}(Y', \hat{Y'}) = \frac{\|Y' - \hat{Y'}\|_2}{\|Y'\|_2},
\end{equation}
where $Y'$ and $\hat{Y'}$ are normalized true and predicted output data, respectively.

We implement the TransDenseUNet architecture and all training experiments using the TensorFlow library (version 2.11) \cite{Tensorflow2015, githubGitHubTensorflowtensorflow} supported by the NumPy library (version 1.24) \cite{Harris-2020-ID6210, githubGitHubNumpynumpy} in the Python programming language (version 3.10.6) \cite{Python2009}. All models are trained on a single Nvidia Tesla P100 16GB GPU, except for TDN528$_\mathrm{30k}$ which is trained on a single Nvidia A100 80GB GPU. Although the training data are prepared using double precision (FP64), the models are trained using single precision (FP32) for significantly faster training sessions and inferences. The training time required for each model can take from several days to a week, depending on the size of the model and the dataset.

%\subsection{Training results and prediction performance}
%\begin{figure*}[htbp]\centering
%\includegraphics[width=1\textwidth]{Figures/loss_and_error_test.png} 
%\caption{Total loss and relative error - PLACE HOLDER.}
%\label{fig:loss_error}
%\end{figure*}
\section{Results and discussion}
\label{sec:results_discussion}

In this section, we validate the training success and evaluate the performance of the trained models in terms of quantitative and qualitative accuracy. Subsequently, we discuss the electric field accuracy calculated from the predicted potential. Furthermore, the need for refinement using a conventional iterative solver to address the spectral bias problem will also be discussed. Finally, we present and briefly discuss the wall time performance of the learned solvers in comparison with a conventional GPU-based iterative solver.

\subsection{Training results}
\begin{figure}[tbp]\centering
\includegraphics[width=0.5\columnwidth]{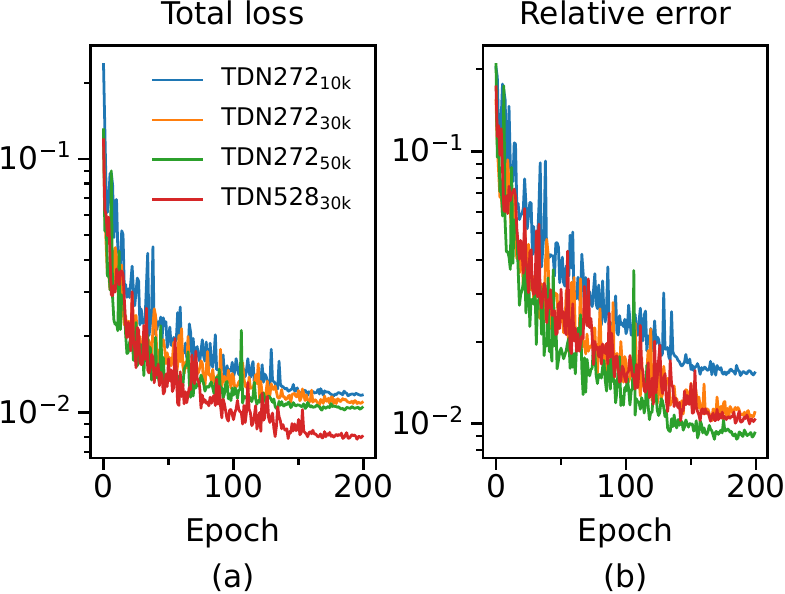} 
\caption{Evolution of (a) validation total loss (equation \eref{eq:l_total})  and (b) relative error (equation \eref{eq:rel_error}) during training.}
\label{fig:val_loss_error}
\end{figure}

Training success can be generally assessed from the evolution of the loss terms on validation data during training. The validation total losses (equation \eref{eq:l_total}) and the relative errors (equation \eref{eq:rel_error}) are shown in Figure \ref{fig:val_loss_error}. It decreases gradually over the training epochs and finally plateaus toward the end of training. There is no sign of overfitting observed in any model, i.e., no increase in validation loss at a later epoch. This indicates good learning convergence. We also observe that after around epoch 125, there are smaller fluctuations in the validation losses. This correlates with a small learning rate ($1.5\cdot 10^{-4}$) at this stage. From this observation, it follows that the use of gradually decreased learning rates implies a stable learning behavior. However, the choice of a small learning rate must always be taken with caution to avoid underfitting and a long training time.

Furthermore, Figure \ref{fig:val_loss_error} shows the effects of the number of training data and resolution on the training. In terms of loss, TDN528$_\mathrm{30k}$ achieves the lowest loss as presented in Figure \ref{fig:val_loss_error}(a). However, this does not reflect directly on the relative error from equation \eref{eq:rel_error}. In Figure \ref{fig:val_loss_error}(b), we can see that TDN272$_\mathrm{50k}$ has the lowest relative error, whereas TDN528$_\mathrm{30k}$ and TDN272$_\mathrm{30k}$ share similar relative error and TDN272$_\mathrm{10k}$ is by far the worst in terms of loss and relative error.

\begin{figure*}[tbp]\centering
\includegraphics[width=1\textwidth]{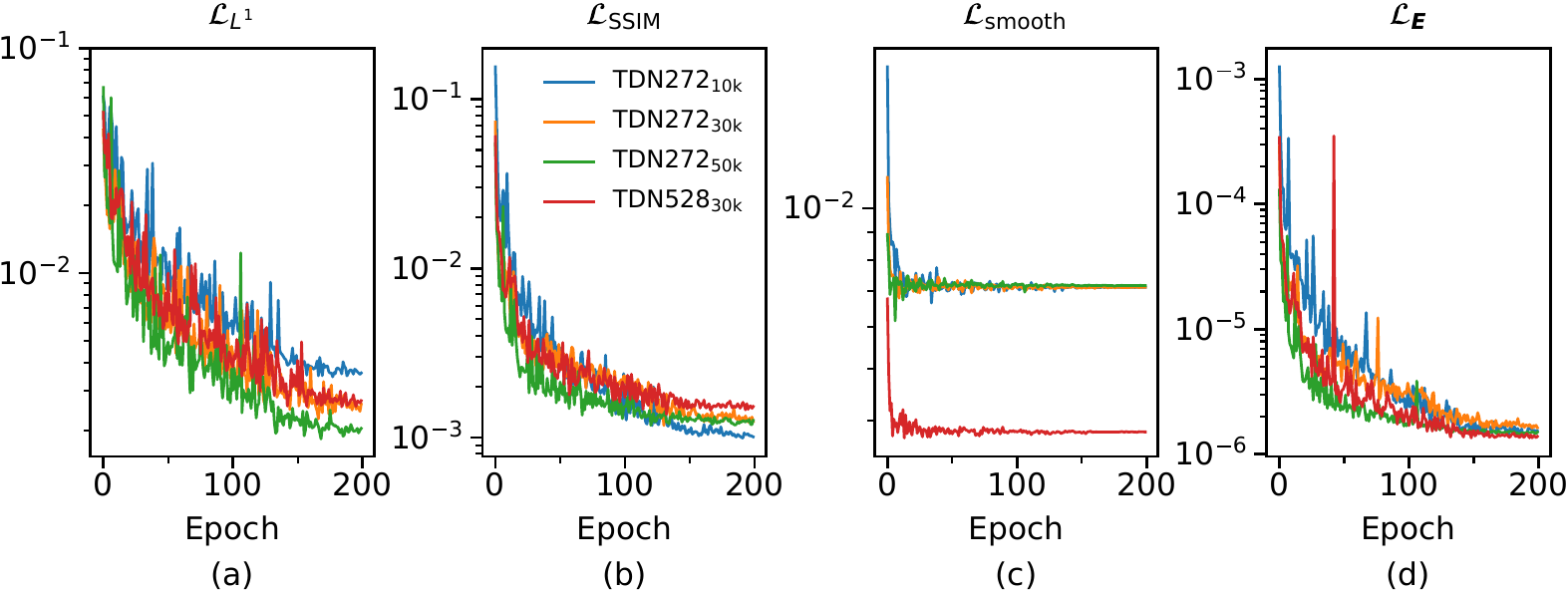} 
\caption{Evolution of individual loss terms during training. From left to right: (a) $L^1$ loss, (b) SSIM loss, (c) disparity smoothness loss, and (d) contextual physics loss terms.}
\label{fig:individual_losses}
\end{figure*}

The loss and relative error discrepancy of TDN528$_\mathrm{30k}$ is mostly attributed to the contributions of individual loss terms as depicted in Figure \ref{fig:individual_losses}. We can see from Figure \ref{fig:individual_losses}(a) that $\mathcal{L}_{L^1}$ follows a very similar trend as observed in the relative error. This suggests $\mathcal{L}_{L^1}$ is the main contributor to the numerical accuracy of the models and that outliers do not contribute significantly. As mentioned in section~\ref{sec:objective_function}, $\mathcal{L}_\mathrm{SSIM}$ and $\mathcal{L}_\mathrm{smooth}$ are mainly intended to enforce smooth and perceptually similar predictions. Figure \ref{fig:individual_losses}(b) implies that the network slightly struggles to learn this property for higher resolutions. To verify this assumption, a hyperparameter tuning for $\mathcal{L}_\mathrm{SSIM}$ is of interest. Nevertheless, the presented SSIM losses still achieved quantitatively satisfying results ($\ll 1$, recall that we set $\lambda = 1$ in $\mathcal{L}_\mathrm{SSIM}$). For $\mathcal{L}_\mathrm{smooth}$, we can see a large discrepancy between the 272 and 528 models as shown in Figure \ref{fig:individual_losses}(c). Notice though that there is no significant contribution from this loss term following the first few epochs. We further assume that removing this loss term may only negligibly affect the overall performance. Lastly, we can see in Figure \ref{fig:individual_losses}(d) that $\mathcal{L}_{\boldsymbol{E}}$ converges similarly for all models at the end of training.

\subsection{Model evaluations}

We evaluate the raw prediction accuracy of the trained models on validation and evaluation sets using the relative $L^2$-norm error $\varepsilon_{L^2}$ from equation \eref{eq:rel_error}. From the results shown in table \ref{table:results}, we can infer that the model trained on the largest dataset, in this case TDN272$_\mathrm{50k}$, outperforms other models. To put things into prespective, TDN272$_\mathrm{50k}$ achieves $\varepsilon_{L^2} = 8.7 \cdot 10^{-3}$ on its validation set, which is better than the performance of the NO model ($\varepsilon_{L^2} = 1.09 \cdot 10^{-2}$) in \cite{Li-2020-ID6192} for a similar yet simpler task, i.e., the Darcy flow problem with homogeneous boundary conditions and a constant source term. It is also worth noting that TDN272$_\mathrm{50k}$ performs competitively with the latest state-of-the-art NO model in \cite{Cao-2021-ID6193} ($\varepsilon_{L^2} = 8.44 \cdot 10^{-3}$ on the Darcy flow problem).

Furthermore, we can see that TDN272$_\mathrm{30k}$ and TDN528$_\mathrm{30k}$, both trained with the same number of training data, perform similarly on their corresponding validation set and test set $D_{*}^{(\mathrm{test})}$. Slight discrepancies are observed on the reference geometry datasets, whereas the TDN528$_\mathrm{30k}$ model performs slightly better than TDN272$_\mathrm{30k}$ on average.

\begin{table*}[tbp]
\caption{\label{table:results} Averaged raw prediction accuracy of the trained models on the validation and evaluation sets. The scores are the prediction relative error $\varepsilon_{L^2}$ from equation \eref{eq:rel_error} with respect to the corresponding ground truth. Here, $*$ means data resolution (which corresponds to the model resolution). The best score from each set is presented in bold.}
\centering \small
\begin{tabular*}{1\textwidth}{@{}l*{15}{@{\extracolsep{0pt plus 12pt}}l}}
\br
Model  & \centre{1}{Validation set}     & \centre{1}{$\mathcal{D}_{*}^{(\mathrm{test})}$} & \centre{1}{$\mathcal{D}_{*}^{(\mathrm{ccrf})}$} & \centre{1}{$\mathcal{D}_{*}^{(\mathrm{dbd1})}$} & \centre{1}{$\mathcal{D}_{*}^{(\mathrm{dbd2})}$}\\
\mr
TDN272$_\mathrm{10k}$ & \centre{1}{$1.527 \cdot 10^{-2}$}         & \centre{1}{$1.584\cdot 10^{-2}$}         & \centre{1}{$3.193\cdot 10^{-2}$}         & \centre{1}{$2.1\cdot 10^{-2}$}          & \centre{1}{$2.229\cdot 10^{-2}$} \\
TDN272$_\mathrm{30k}$ & \centre{1}{$1.086\cdot 10^{-2}$}         & \centre{1}{$1.083\cdot 10^{-2}$}         & \centre{1}{$1.91\cdot 10^{-2}$}         & \centre{1}{$1.014\cdot 10^{-2}$}          & \centre{1}{$1.893\cdot 10^{-2}$}  \\
TDN272$_\mathrm{50k}$ & \centre{1}{$\mathbf{8.67\cdot 10^{-3}}$}& \centre{1}{$\mathbf{9.51\cdot 10^{-3}}$}& \centre{1}{$8.73\cdot 10^{-3}$}         & \centre{1}{$\mathbf{5.29\cdot 10^{-3}}$} & \centre{1}{$\mathbf{8.81\cdot 10^{-3}}$}\\
TDN528$_\mathrm{30k}$ & \centre{1}{$1.084\cdot 10^{-2}$}         & \centre{1}{$1.127\cdot 10^{-2}$}         & \centre{1}{$\mathbf{7.79\cdot 10^{-3}}$}& \centre{1}{$9.95\cdot 10^{-3}$}          & \centre{1}{$2.1\cdot 10^{-2}$}  \\
\br
\end{tabular*}
\end{table*}

%\begin{table*}
%\caption{\label{table:results} Averaged one-shot prediction performance of the trained models (measured using relative error $\varepsilon_{L^2}$ excluding the paddings) on the validation and evaluation sets. Here, $\cdot$ means the resolution of the data (without padding) corresponding to the model resolution. Scores are presented in $(\times 10^2)$ and best score from each set is presented in bold.}
%\centering \small
%\begin{tabular*}{1\textwidth}{@{}l*{15}{@{\extracolsep{0pt plus 12pt}}l}}
%\br
%       & \centre{2}{Val. set}     & \centre{2}{$\mathcal{D}_{*}^{(\mathrm{test})}$} & \centre{2}{$\mathcal{D}_{*}^{(\mathrm{ccrf})}$} & \centre{2}{$\mathcal{D}_{*}^{(\mathrm{dbd1})}$} & \centre{2}{$\mathcal{D}_{*}^{(\mathrm{dbd2})}$}\\
%\mr
%Models               & Raw          & Refined & Raw & Refined   & Raw & Refined         & Raw & Refined   & Raw & Refined  \\
%\mr
%TransDUNe272$_\mathrm{10k}$ & 0.0153          & n/a  & 0.0158          & n/a & 0.0319          & n/a & 0.021           & n/a & 0.0223          & n/a\\
%TransDUNe272$_\mathrm{30k}$ & 0.0109          & n/a  & 0.0108          & n/a & 0.0191          & n/a & 0.0101          & n/a & 0.0189          & n/a\\
%TransDUNe272$_\mathrm{50k}$ & \textbf{0.0087} & n/a  & \textbf{0.0095} & n/a & 0.0087          & n/a & \textbf{0.0053} & n/a & \textbf{0.0088} & n/a\\
%TransDUNe528$_\mathrm{30k}$ & 0.0108          & n/a  & 0.0113          & n/a & \textbf{0.0078} & n/a & 0.00995         & n/a & 0.021  & n/a\\
%\br
%\end{tabular*}
%\end{table*}

\begin{figure*}[tbp]\centering
\includegraphics[width=1\textwidth]{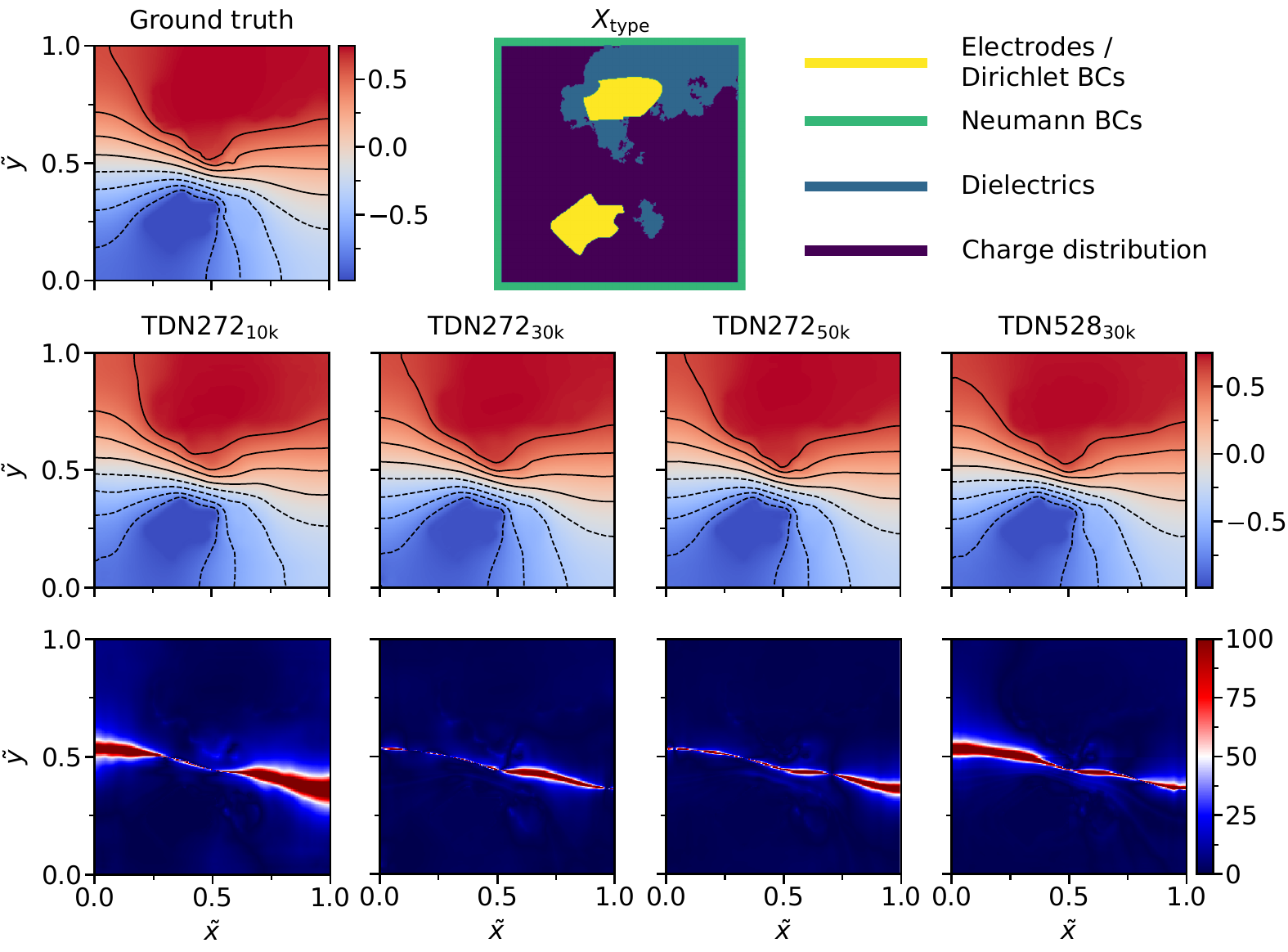} 
\caption{Example of a predicted potential $\tilde{\phi}$ (denormalized and excluding paddings) from the trained models (middle panels) on randomly generated input data and their corresponding error maps in percentage (bottom panels) with respect to the ground truth (left most top panel). From left to right (of middle panels): predicted potential from TDN272$_\mathrm{10k}$ ($\varepsilon_{L^2}= 2.5 \cdot 10^{-2}$), predicted potential from TDN272$_\mathrm{30k}$ ($\varepsilon_{L^2}=1.202 \cdot 10^{-2}$), predicted potential TDN272$_\mathrm{50k}$ ($\varepsilon_{L^2}=8.98 \cdot 10^{-3}$), and predicted potential from TDN528$_\mathrm{30k}$ ($\varepsilon_{L^2}=1.673 \cdot 10^{-2}$).}
\label{fig:raw_prediction}
\end{figure*}

% Caption for Fig_predictions.pdf
%\caption{Example of predicted potential $\tilde{\phi}$ \unsure{(denormalized and excluding paddings)} from the trained models (top panels) on randomly generated input data and their corresponding error maps in percentage (bottom panels) \unsure{with respect to the ground truth}. From left to right: ground truth potential, predicted potential from TDN272$_\mathrm{10k}$ ($\varepsilon_{L^2}$=0.025), predicted potential from TDN272$_\mathrm{30k}$ ($\varepsilon_{L^2}$=0.01202), predicted potential TDN272$_\mathrm{50k}$ ($\varepsilon_{L^2}$=0.00898), and predicted potential from TDN528$_\mathrm{30k}$ ($\varepsilon_{L^2}$=0.01673).}

Figure \ref{fig:raw_prediction} shows an example of raw prediction from each model for the same random reactor geometry data sampled from $\mathcal{D}^{\mathrm{(test)}}_{*}$. We can see from the middle panels of the figure that all models produce qualitatively similar predictions to the ground truth (left most top panel). A closer inspection reveals that TDN272$_\mathrm{50k}$ produces a better prediction, which can be verified by the accuracy of the contour lines. This is also well reflected in the relative error, which is the smallest ($\varepsilon_{L^2}=8.98 \cdot 10^{-3}$). The bottom panels of Figure \ref{fig:raw_prediction} show the error maps of the predictions in percentage (the values are truncated at 100). We can see that most of the errors accumulate at the zero crossing, i.e., the value transition from positive to negative (or vice versa). This is mainly due to the division by close to zero (i.e., very small numbers) and the models' inability to predict values with great precision, which is a trade-off of using the FP32 format (i.e., finite precision and round-off errors).

\subsection{Accuracy of the predicted electric field and prediction refinement}
\label{sec:accuracy_efield}
With regard to regression tasks and given the complexity of the problem, TDN272$_\mathrm{50k}$ performs exceedingly well both qualitatively and quantitatively. However, predicting or calculating the potential profile is often only halfway through the field calculation step in LTP simulations. The next and final step is to calculate the electric field, and in this case, we use the predicted potential given by $\tilde{\boldsymbol{E}}=-\nabla \tilde{\phi}$.

\begin{figure*}[tbp]\centering
\includegraphics[width=1\textwidth]{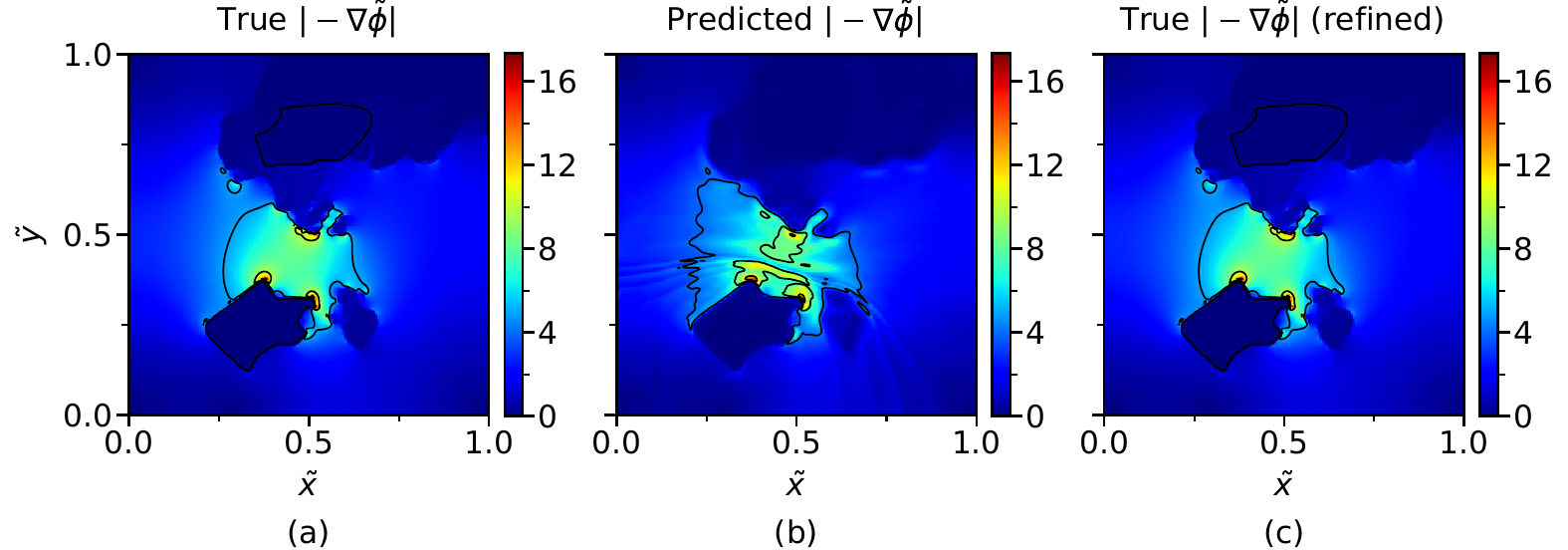} 
\caption{Comparison of (a) true, (b) predicted, and (c) refined absolute electric fields. The true and predicted absolute electric fields are calculated from the ground truth in Figure \ref{fig:raw_prediction} and the predicted potential profile from the TDN272$_\mathrm{50k}$ in Figure \ref{fig:raw_prediction}, respectively. The predicted absolute electric field is calculated from the refined predicted potential using 10 iterations of a GPU-accelerated GMRES solver.}
\label{fig:randomgeo_efield}
\end{figure*}

\begin{figure}[tbp]\centering
\includegraphics[width=0.5\columnwidth]{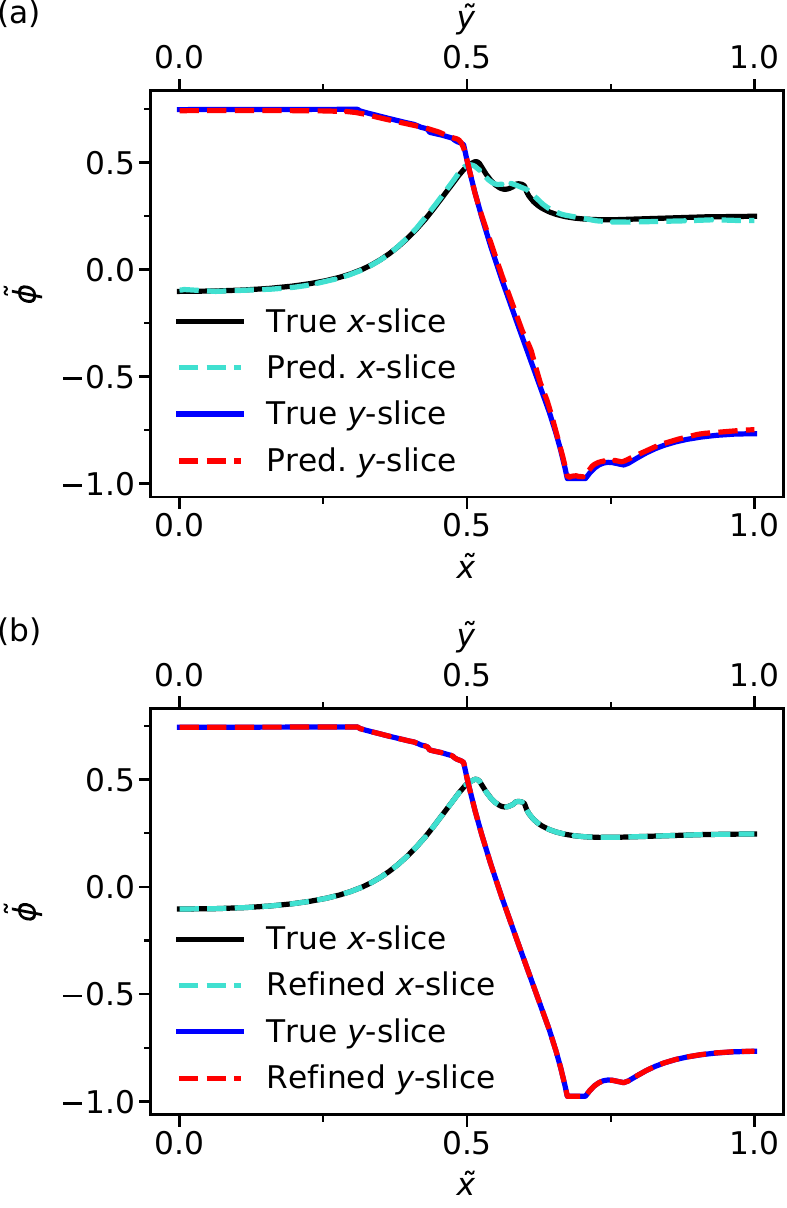} 
\caption{1D line plots of (a) predicted and (b) refined potential profiles against the corresponding true profiles across the $x$ and the $y$ axes (taken from the middle of each axis). The true and predicted potential profiles are based on the ground truth and the prediction of TDN272$_\mathrm{50k}$ in Figure \ref{fig:raw_prediction}, respectively.}
\label{fig:sliceplots}
\end{figure}

Figure \ref{fig:randomgeo_efield}(a) and Figure \ref{fig:randomgeo_efield}(b) depict the absolute electric fields calculated from the ground truth potential profile and from the potential profile predicted by TDN272$_\mathrm{50k}$ (middle panels of Figure \ref{fig:raw_prediction}), respectively.  It is clear from the figures that the calculated electric field from the prediction is both qualitatively and quantitatively dissimilar to the ground truth. This is because small fluctuations in the potential profile $\tilde{\phi}$ can affect the gradient of the electric field greatly. We can see these small fluctuations better in the 1D line plots of the potential (taken along the $x$ and $y$ directions through the center of the domain, respectively) as presented in Figure \ref{fig:sliceplots}(a). Although the profiles match closely, small fluctuations can still be observed, especially in the regions with high spatial dynamics corresponding to the high-frequency features of the potential profile. Further observation on Figure \ref{fig:randomgeo_efield}(b) reveals that the model struggles to recover the high-frequency features, i.e., spatial details mostly found at the interface between different materials.

Neural networks have been observed to initially learn the low-frequency features of the target function, and very slowly converge on the high-frequency features; this phenomenon is referred to as the spectral bias \cite{Markidis-2021-ID6224,Rahaman-2018-ID6228,Cao-2019-ID6229,Xu-2019-ID6230,Xu-2019-ID6231}. Unfortunately, the training time required for the networks to learn the high-frequency features properly may not be tractable for practical use cases. To mitigate this problem, we follow the same method as demonstrated in \cite{Xu-2019-ID6231,Ozbay-2021-ID6201,Markidis-2021-ID6224}. That is to refine the predicted $\tilde{\phi}$ using a conventional iterative solver, i.e., by using the predicted $\tilde{\phi}$ as an initial guess for the conventional solver instead of a zero initial guess. To this end, we utilize a GPU-accelerated GMRES solver from the AMGX library\cite{Naumov-2015-ID6232,githubGitHubNVIDIAAMGX} (wrapped in the PyAMGX library\cite{githubGitHubShwinapyamgx} for usage within Python) and refine the prediction for 10 iterations. The GMRES solver is preconditioned using an Algebraic Multigrid (AMG) solver from the same library \cite{Naumov-2015-ID6232}. We set the absolute tolerance for stopping criteria to $10^{-10}$ for all cases in this work.

Figure \ref{fig:randomgeo_efield}(c) shows the electric field distribution calculated from the refined prediction. As apparent from a comparison with Figure \ref{fig:randomgeo_efield}(b), the refinement alleviates the high-frequency problem and produces a qualitatively identical electric field to the ground truth (Figure \ref{fig:randomgeo_efield}(a)). We also observe a considerable decrease in relative error, which reduces from $\varepsilon_{L^2} = 8.98 \cdot 10^{-3}$ to $\varepsilon_{L^2} = 3.7 \cdot 10^{-4}$. Figure \ref{fig:sliceplots}(b) further quantifies this improvement, where the predicted and refined line profiles agree with the true profiles. It is worth noting that for this particular example, the iterative solver with a zero initial guess achieves $\varepsilon_{L^2} = 1.034 \cdot 10^{-2}$ after 10 iterations, and the resulting potential profile remains qualitatively suboptimal (see Figure \ref{fig:refinevolutions} in \ref{appendix:refinevolutions}). Furthermore, the iterative solver with a zero initial guess requires 68 iterations until convergence (with tolerance = $10^{-10}$), whereas the solver with a prediction from TDN272$_\mathrm{50k}$ as an initial guess requires 57 iterations, which is about 16\% fewer. Using this strategy of combining a neural network-based solver (such as TDN272$_\mathrm{50k}$) and a conventional solver, one can arrive at a reasonable solution of the Poisson equation with fewer iterations and thus potentially faster calculation time. It is worth noting that a highly accurate initial prediction may result in a significantly faster computation time (fewer iterations needed for refinement) and a more accurate final solution. Therefore, it is imperative to address the problem of spectral bias in future developments.

\begin{figure*}[tbp]\centering
\includegraphics[width=1\textwidth]{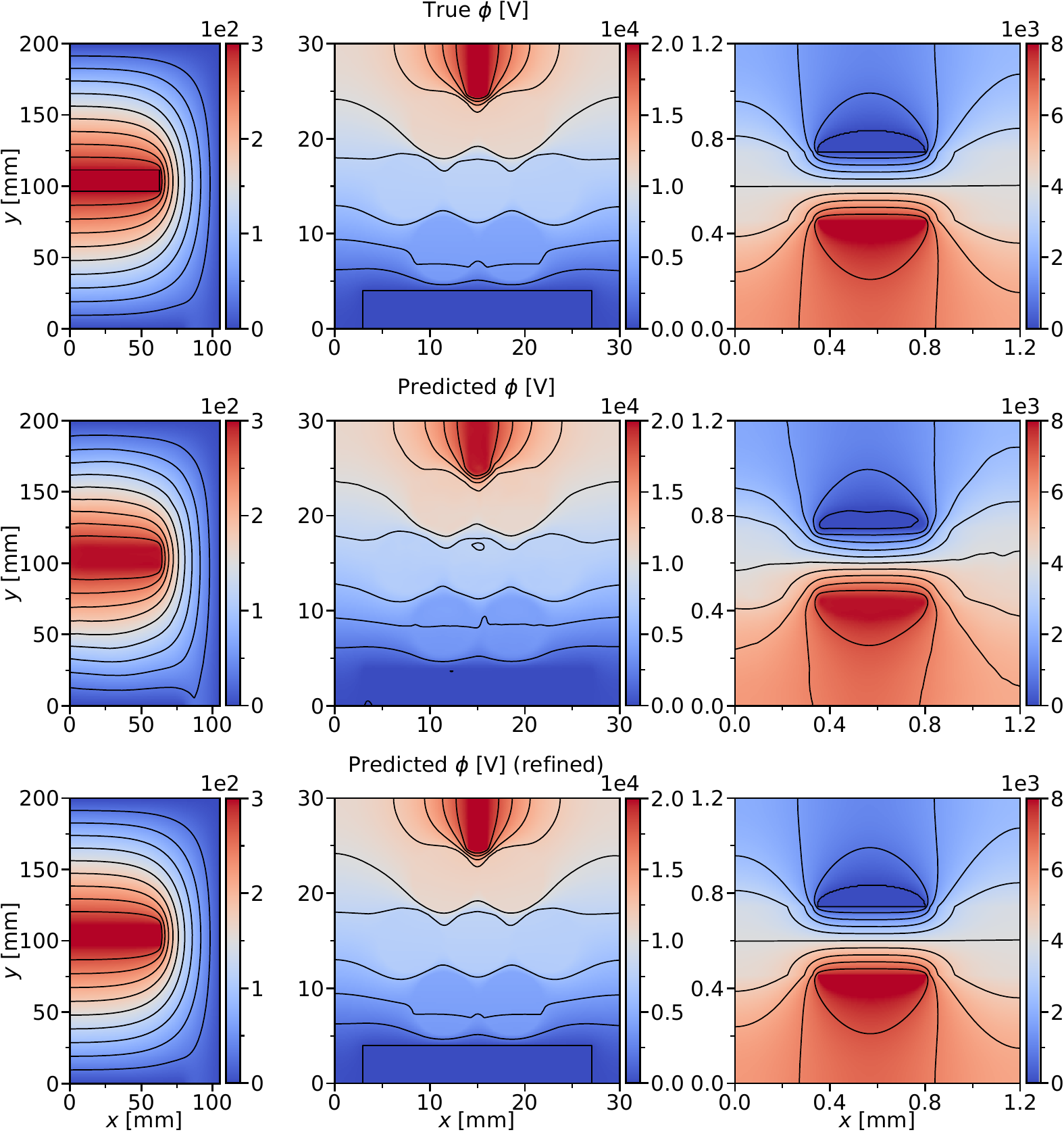} 
\caption{Raw and refined potential profile predictions of reference reactor geometry data in comparison with the ground truths. From left to right: asymmetric CCRF discharge, packed-bed DBD, and surface DBD reactor geometries. From top to bottom: true, predicted, and refined potential profiles. From left to right, the relative errors $\varepsilon_{L^2}$ of the raw predictions (middle panels) are $1.2 \cdot 10^{-2}$, $5.5 \cdot 10^{-3}$, and $9.4 \cdot 10^{-3}$. For the refined predictions (bottom panels), the relative errors are 4.845$\cdot 10^{-7}$, $1.04 \cdot 10^{-3}$, and $2.9 \cdot 10^{-4}$, following the same order.}
\label{fig:known_geo_potential}
\end{figure*}

\begin{figure*}[htbp]\centering
\includegraphics[width=1\textwidth]{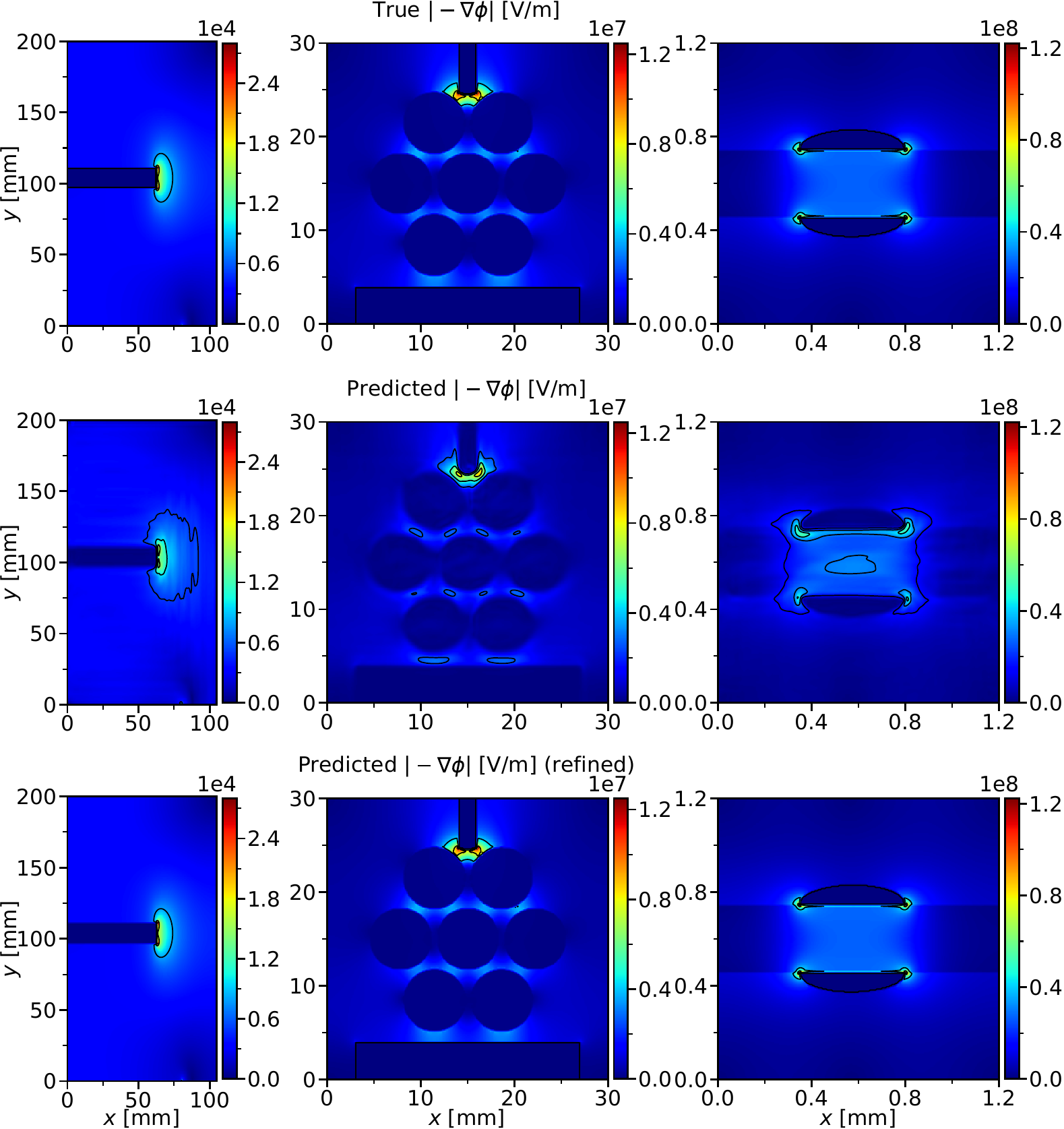} 
\caption{Raw and refined absolute electric fields of reference reactor geometry data in comparison with the ground truths. From left to right: asymmetric CCRF discharge, packed-bed DBD, and surface DBD reactor geometries. From top to bottom: true, predicted, and refined absolute electric fields.}
\label{fig:known_geo_efield}
\end{figure*}

At this point, the ability of TDN272$_\mathrm{50k}$ to generalize on geometries excluded during the training should be discussed. Figure \ref{fig:known_geo_potential} compares the results for the true (top panels) and predicted (middle panels) potential profiles of reference reactor geometry data (presented in physical units). The agreement suggests that the strategy of training the networks with randomized geometry data is effective in ensuring the generalizability of the model to various geometry configurations. The bottom panels of Figure \ref{fig:known_geo_potential} show the refined potential profile predictions, where we can observe the apparent corrections in the contour lines. The final result is exceedingly similar to the ground truths. In terms of quantitative comparison, the relative errors of the predicted potential profiles amount to $\varepsilon_{L^2}=$ $1.2 \cdot 10^{-2}$, $5.5 \cdot 10^{-3}$, and $9.4 \cdot 10^{-3}$ for asymmetric CCRF discharge, packed-bed DBD, and surface DBD reactors, respectively. This is more or less consistent with the expectations for the test dataset $\mathcal{D}^\mathrm{(test)}_{*}$ in table \ref{table:results}. However, this calls for a similar need for refinement for the electric field calculation. This can be clearly observed in the electric fields as shown in Figure \ref{fig:known_geo_efield} (middle and bottom panels), which again exhibits the problem of a spectral bias. 

The relative errors of the refined potential profiles (middle panels of Figure \ref{fig:known_geo_potential}), following the same order, are $\varepsilon_{L^2}=$ $4.845\cdot 10^{-7}$, $1.04 \cdot 10^{-3}$, and $2.9 \cdot 10^{-4}$. This reveals that the existence of dielectric materials and the geometry complexity impact the overall number of iterations needed to achieve satisfying results. For instance, the asymmetric CCRF discharge reactor geometry, the simplest geometry, has the lowest error after 10 iterations of refinement. Whereas the packed-bed DBD reactor geometry, the most complex geometry, has the highest error of the three after the same number of iterations.

\subsection{Inference times of the learned solvers}
Although the computation speedup of the learned solvers is not the primary focus of the present work, it is still worth discussing their wall time performance. Modern machine learning libraries are well-optimized for GPUs. While the training of the models can take several days, the inference times of the trained models only take a fraction of a second on GPUs. For example, TDN272$_\mathrm{50k}$ requires 70\,ms on average for one inference and  124\,ms for TDN528$_\mathrm{30k}$ on a single Nvidia RTX 3080 12GB GPU. In comparison, a conventional iterative solver (GMRES) \cite{Naumov-2015-ID6232} running on the same GPU requires around 333\,ms and 503\,ms on average to calculate the solutions in $256\times 256$ and $512 \times 512$ domain resolutions, respectively. The learned solvers are in fact faster than the conventional solver. Additionally, the inference times of the learned solvers scale nonlinearly with a batch prediction. For instance, for a batch of 8 predictions, the total inference times for TDN272$_\mathrm{50k}$ and TDN528$_\mathrm{30k}$ are 155\,ms and 683\,ms, respectively. Typically, the inference time stays the same for any input data as it depends solely on the sizes of the model and input data. The fast and stable inference time (equivalent to solving time) is indeed an attractive feature of a learned solver.

As discussed in section~\ref{sec:accuracy_efield}, using a learned solver alone is not enough to achieve reasonable numerical accuracy for practical use cases, and a prediction refinement must be conducted using a conventional iterative solver. We show earlier in section~\ref{sec:accuracy_efield} that using a predicted potential from a learned solver can decrease the overall iteration steps of a conventional iterative solver. And greater prediction accuracy results in fewer iteration steps, thus faster calculation time. However, further investigation is necessary to determine the quantitative computation speedup that the learned solvers can provide. Nonetheless, this work demonstrates the potential of the learned solvers to enhance conventional iterative solvers in achieving faster calculation times.

\section{Conclusion and outlook}
\label{sec:conclusion_outlook}
In this paper, we present a machine learning-based 2D Poisson solver for use cases in low-temperature plasma simulations that can accommodate varying degrees of geometric complexity and various boundary condition configurations (Dirichlet, Neumann, and mixed boundary conditions). The followings are the summary and highlights of the used approach:
\begin{itemize}
    \item Hybrid architecture consisting of CNNs and transformer networks called TransDenseUNet (TDN), which is based on the existing TransUNet architecture developed initially for medical image segmentation tasks.
    \item Weighted multiterm loss function consisting of $L^1$-norm loss term ($\mathcal{L}_{L^1}$), loss terms adopted from computer vision such as SSIM and disparity smooth loss terms ($\mathcal{L}_\mathrm{SSIM}$ and $\mathcal{L}_\mathrm{smooth}$), and finally, a contextual-physics loss term ($\mathcal{L}_{\boldsymbol{E}}$), which in this case takes the form of the relative error of the gradient of the predicted potential (electric field).
    \item Using the dimensionless Poisson's equation as a mean of feature scaling.
    \item Highly-randomized data generation scheme: random reactor geometries (shapes of electrodes and dielectric materials), charge distributions, and boundary condition configurations. 
    \item Using paddings to improve the contributions of outer boundary conditions in the input data.
    \item Multichannel input data configuration that exploits co-location properties of the input parameters encoded in several channels: type, value, and (two) position channels.
    \item Prediction refinement using a conventional iterative solver on a GPU.
\end{itemize}

From the presented results, the best performing learned solver (TDN272$_\mathrm{50k}$) is shown to generalize well on new reactor geometries with practically identical (or better) accuracy to the one obtained from the validation and test sets. We also show that better solver accuracy can be attained by increasing the amount of training data. Additionally, increasing the resolution (of both data and model) has little to no effect on the accuracy. The results also qualitatively show that the predicted potential profiles exhibit an exceedingly good level of smoothness.

With regard to the practical application of the developed learned solver, raw solutions (predictions) from the learned solver, although very accurate as far as regression tasks are concerned, are not yet accurate enough to be used directly in LTP simulations. Therefore, we recommend conducting a prediction refinement using a conventional iterative solver. We show that this approach can achieve a numerically accurate and potentially faster solution (fewer iteration steps until convergence) than a conventional solver alone running on a GPU for high-resolution domains.

This work reveals that the learned solvers suffer from the spectral bias problem, wherein the model is unable to effectively learn the high-frequency features (primarily present at the interfaces between distinct materials within the domain) within a reasonable training time. Therefore, it is highly imperative to tackle this problem in future learned solvers. We presume further theoretical work on the network architecture and optimization technique is needed to resolve the spectral bias problem. Presently, this work has addressed problems defined on structured Cartesian grids. In many cases, plasma simulations employ unstructured grids or meshes for more stable and accurate simulations of certain phenomena within plasma physics. Hence, it is also of great interest to consider this requirement in future developments. Preliminary work has been done by Pfaff \textit{et al.} \cite{Pfaff-2020-ID6233} and  L\"otzsch \textit{et al.} \cite{Lotzsch-2022-ID6234} that addresses this requirement using graph neural networks (GNNs). In the immediate next steps, we will conduct model optimization and evaluate the developed (and optimized) learned solvers in real simulation settings such as Particle-in-Cell/Monte Carlo Collision (PIC/MCC) or fluid-Poisson simulations. The evaluation will address aspects such as the accuracy of the simulations using the learned solvers in place of pure conventional solvers, and how much practical speedup the learned solvers can provide.

As final remarks, the generalization capability of the learned solver can prevent the  high cost of computation incurred from retraining. And it is worth bearing in mind that the proposed approach can be straightforwardly adapted to many different 2D PDE and image-to-image translation problems. Lastly, a pure machine-learned Poisson solver that tackles all the requirements presented in this work may not be readily available in the short term, and a marriage between ML-based and conventional solvers is recommended as far as practical applications are concerned. The proposed approach is the next step further in achieving a generic and high-performing machine learning-based Poisson solver, especially for LTP simulations in complex geometries.

\section*{Acknowledgment}
J.T. acknowledges funding by the Deutsche Forschungsgemeinschaft (DFG, German Research Foundation) -- Project-ID 434434223 -- SFB 1461. He was also partially funded by Tokyo Electron Technology Solutions Ltd. I.C.S. would like to thank Igor Semenov and Aleksandar P. Jovanović for fruitful discussions.

\printbibliography
\section*{References}
%\bibliographystyle{iopart-num}
%\bibliography{refs,refs-mod}

\appendix
\section{Structural Similarity Index Measure (SSIM)}
\label{appendix:ssim}
Unlike other metrics such as mean absolute error (MAE) and mean squared error (MSE), which consider the absolute difference between two images, SSIM considers the difference between two images based on the perceived change in the local information (local statistics). The original publication by Wang \textit{et al.} \cite{Wang-2004-ID6212} determines this local information into three components: luminance, contrast, and structural given by (in the same order)
\begin{eqnarray}
l(x,y)=\frac{2\mu_x\mu_y+c_1}{2\mu_x^2+\mu_y^2 + c_1}, \\
c(x,y)=\frac{2\sigma_x\sigma_y+c_2}{2\sigma_x^2+\sigma_y^2+c_2},\\
s(x,y) = \frac{\sigma_{xy}+c_3}{\sigma_x \sigma_y + c_3},
\end{eqnarray}
where
\begin{itemize}
    \item $\mu_x$ is the mean of a local area in image X,
    \item $\mu_y$ is the mean of a local area in image Y,
    \item $\sigma_x$ is the standard deviation of a local area in image X,
    \item $\sigma_y$ is the standard deviation of a local area in image Y,
    \item $\sigma_{xy}$ is the co-variance of a local area between image X and Y,
    \item $x$ and $y$ are a local area in the image X and Y, respectively, determined by an $n\times n$ Gaussian window,
    \item $c_1 = (k_1 \lambda)^2$, $c_2=(k_2 \lambda)^2$, and $c_3 = c_2/2$ are constants,
    \item $\lambda$ is the dynamic range of the input images (difference between minimum and maximum values),
    \item lastly, $k_1=0.01$ and $k_2=0.03$ following the original publication.
\end{itemize}
The local statistics are computed using a Gaussian window which slides over the measured images (X and Y). Finally, the similarity of the two images is a scalar value given by
\begin{eqnarray}
    \mathrm{SSIM}(X,Y) &= \frac{1}{N} \sum_{i=1}^N [l(x_i,y_i) \cdot c(x_i,y_i) \cdot s(x_i,y_i)] \nonumber\\
    &= \frac{1}{N} \sum_{i=1}^N \frac{(2\mu_{x_i} \mu_{y_i} + c_1)(2\sigma_{xy_i}+c_2)}{(\mu_{x_i}^2+ \mu_{y_i}^2+c_2)(\sigma_{x_i}^2+\sigma_{y_i}^2+c_2)},
\end{eqnarray}
where
\begin{itemize}
    \item $l(x_i,y_i)$, $c(x_i,y_i)$, and $s(x_i,y_i)$ are the luminance, contrast, and structural of the $i$-th local image area, respectively,
    \item $\mu_{x_i}$ is the $i$-th mean of local area in image X,
    \item $\mu_{y_i}$ is the $i$-th mean of local area in image Y,
    \item $\sigma_{x_i}$ is the $i$-th standard deviation of a local area in image X,
    \item $\sigma_{y_i}$ is the $i$-th standard deviation of a local area in image Y,
    \item $\sigma_{xy_i}$ is the $i$-th co-variance of a local area between image X and Y,
    \item $N$ is the number of local image areas processed by the Gaussian window.
\end{itemize}
Note that the presented formulation of SSIM is for single-channel images. In practice, for multichannel images, one can compute the SSIM of each channel and take the average of all channels. For a more detailed description of SSIM, we refer interested readers to the original publication by Wang \textit{et al.} \cite{Wang-2004-ID6212}. 

%\newpage

\section{}
\label{appendix:refinevolutions}

\begin{figure*}[htbp]\centering
\includegraphics[width=1\textwidth]{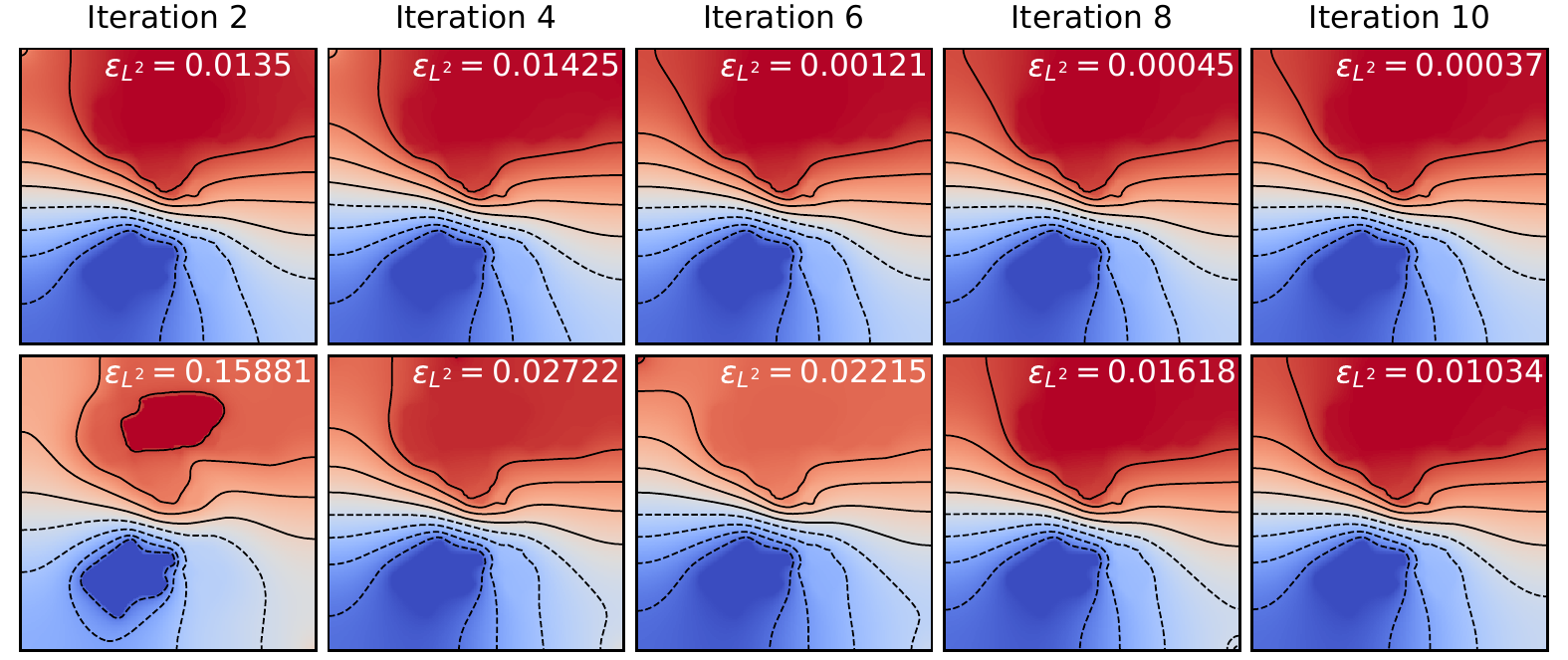} 
\caption{Evolution of iterative solutions. The top panels are the solution evolution of an iterative solver using a prediction from TDN272$_\mathrm{50k}$ as an initial guess (rTDN$_\mathrm{GPU}$), and the bottom panels are from an iterative solver using a zero initial guess (GMRES$_\mathrm{GPU}$). Note that at iteration 6, rTDN$_\mathrm{GPU}$ has already achieved significantly better accuracy than GMRES$_\mathrm{GPU}$ at iteration 10 and produced a qualitatively identical solution to the ground truth presented in Figure \ref{fig:raw_prediction}.}
\label{fig:refinevolutions}
\end{figure*}

\end{document}